\documentclass{article}
\usepackage{natbib}
\usepackage{amsmath}
\numberwithin{equation}{section} %Formulas are numbered by section

\usepackage{indentfirst}
\usepackage{subfigure} %Insert multiple  images simultaneously
\usepackage{graphicx}
\usepackage{float}
\usepackage{caption}
\usepackage[top=1in, bottom=1in, left=1.25in, right=1.25in]{geometry}
\usepackage[justification=centering]{caption}
\usepackage[normalem]{ulem}
\usepackage[toc,page,title,titletoc,header]{appendix}
\usepackage{multirow}
\usepackage{algorithm,algorithmic}

\usepackage{booktabs}
\usepackage{mathtools}
\usepackage{rotating}
\usepackage[colorlinks=true,linkcolor=blue,urlcolor=black,bookmarksopen=true]{hyperref}
\usepackage{bookmark}
\usepackage[titletoc]{appendix}   % package for adding appendix
\usepackage{soul}    % package for strikethrough font
\usepackage{setspace}
\usepackage{numprint}
\usepackage{bm}
\usepackage{tablefootnote}
\usepackage{threeparttable}
\usepackage{hyperref}
\usepackage{amssymb}

\npdecimalsign{.}
\nprounddigits{3}
\hypersetup{
colorlinks,
citecolor=[RGB]{0,0,0},
linkcolor=[RGB]{0,0,0},
urlcolor=black}
\makeatletter

\newcommand{\Rmnum}[1]{\expandafter\@slowromancap\romannumeral #1@}
\makeatother
\usepackage{amsthm}
\makeatletter
\theoremstyle{plain}
\newtheorem{assumption}{\protect\assumptionname}
\theoremstyle{plain}
\newtheorem{thm}{\protect\theoremname}[section]
\theoremstyle{plain}
\newtheorem{lem}{\protect\lemmaname}[section]
\numberwithin{equation}{section}

\providecommand{\assumptionname}{Assumption}
\providecommand{\lemmaname}{Lemma}
\providecommand{\theoremname}{Theorem}

\makeatother
\usepackage{latexsym}
\begin{document}
\title{\textbf{Endogenous Treatment Effect Estimation with some Invalid and Irrelevant Instruments}   }
\author{ Qingliang Fan \thanks{Department of Economics, The Chinese University of Hong Kong. E-mail: michaelqfan@gmail.com.}\and Yaqian Wu \thanks{School of Economics, Xiamen University. E-mail: yaqian2018@stu.xmu.edu.cn.}}

\maketitle

\begin{abstract}
\doublespacing
  Instrumental variables (IV) regression is a popular method for the estimation of the endogenous treatment effects. Conventional IV methods require all the instruments are relevant and valid. However, this is impractical especially in high-dimensional models when we consider a large set of candidate IVs. In this paper, we propose an IV estimator robust to the existence of both the invalid and irrelevant instruments (called R2IVE) for the estimation of endogenous treatment effects. This paper extends the scope of \emph{sisVIVE} \citep{Kang2015Instrumental} by considering a true high-dimensional IV model and a nonparametric reduced form equation. It is shown that our procedure can select the relevant and valid instruments consistently and the proposed R2IVE is root-$n$ consistent and asymptotically normal. Monte Carlo simulations demonstrate that the R2IVE performs favorably compared to the existing high-dimensional IV estimators (such as, NAIVE \citep{Fan2018Nonparametric} and \emph{sisVIVE} \citep{Kang2015Instrumental}) when invalid instruments exist. In the empirical study, we revisit the classic question of trade and growth \citep[]{Frankel1999Trade}.
\end{abstract}
\newpage
\doublespacing
\section{Introduction}\label{intro}
Instrumental variables (IV) method is useful for the identification of the endogenous treatment effects in empirical studies. The internal validity of IV analysis requires that instruments are associated with the treatment (A1: Relevance Condition), have no direct pathway to the outcome (A2: Exclusion Condition), and are not related to unmeasured variables that affect the treatment and the outcome (A3: Exogenous Condition). Finding the instruments that satisfy assumptions (A1)-(A3) is often quite a challenging task for empirical researchers. The inclusion of many redundant instruments (violation of A1) provides poor finite sample properties of estimators. Two-stage least squares (2SLS) tend to have large biases when many weak instruments are used \citep{Bekker1994Alternative}. An instrument is deemed invalid, if it has a direct effect on the outcome (violation of  A2), or an indirect association with the outcome through unobserved confounders (violation of  A3). Using an invalid instrument will lead to inconsistency of the 2SLS and LIML estimators \citep{Kolesar2015Identification}. It thus motivated this study to develop a method that could select relevant and valid instruments from a large set of candidate instruments.

There is a strand of literature on instruments selection under the exclusion condition and exogenous condition. In a seminal work on IV selection, \cite{Donald2001Choosing} consider a procedure that minimizes higher-order asymptotic MSE which relies on a priori knowledge of the order of instruments. \cite{Kuersteiner2010} use model averaging to construct optimal instruments. \cite{Okui2011} considers ridge regression for estimating the first-stage regression. \ In high-dimensional setting, \cite{Gautier2011} propose an estimation method related to the Dantzig selector and the square-root Lasso when the structural equation in an instrumental variables model is itself very high-dimensional. \cite{Belloni2012sparse}  introduce the Post Lasso method which extends IV estimation to high dimension with heteroscedastic and non-Gaussian random disturbances. \cite{Caner2015Hybrid}  use the adaptive Lasso estimator to eliminate the irrelevant instruments.  \cite{Lin2015Regularization} propose a two-stage regularization framework for identifying and estimating important covariates' effects while selecting and estimating optimal instruments where the dimensions of covariates and instruments can both be much larger than the sample size. \cite{Fan2018Nonparametric} study a nonparametric approach regarding a general nonlinear reduced form equation to achieve a better approximation of the optimal instruments with the adaptive group Lasso.

In the invalid IV settings, \cite{liao2013} proposes an adaptive GMM shrinkage estimator to select valid moment consistently.  Also in GMM framework, \cite{Xu2015Select} use an information-based adaptive GMM shrinkage estimation to select both the relevant and valid moments consistently and \cite{Caner2018invalid} develop the adaptive Elastic-Net generalized method of moments (GMM) estimator in large dimensional models with potentially (locally) invalid moment conditions. The aforementioned  papers need a prior knowledge about a subset of instruments that is known to be valid and contains sufficient information for identification and estimation of the causal effects. In contrast, \cite{Kang2015Instrumental} propose a Lasso type procedure to identify and select the set of invalid instruments without any prior knowledge about which instruments are potentially valid or invalid. \cite{Windmeijer2018} suggest a consistent median estimator using adaptive Lasso which obtains the oracle properties. \cite{Guo2018} propose a Two-Stage Hard Thresholding (TSHT) with voting procedure that selects relevant and valid instruments and produces valid confidence intervals for the causal effect, which also works in the high dimension setting. 
% \cite{Bietal2020} discuss inferring treatment effects after testing instrument strength in linear models. 

In this paper, we develop a method that extends the scope of \emph{sisVIVE} \citep{Kang2015Instrumental} by considering a true high-dimensional instrumental variable setting. Without knowing which instrument is irrelevant or invalid, we consider a large set of candidate instruments $\mathcal{IV}=\mathcal{IV}_{1}\cup\mathcal{IV}_{2}\cup\mathcal{IV}_{3}\cup\mathcal{IV}_{4}$, where $\mathcal{IV}_{1}$ denotes the set of relevant and valid instruments, $\mathcal{IV}_{2}$ denotes the set of relevant and invalid instruments, $\mathcal{IV}_{3}$ denotes the set of irrelevant and valid instruments and $\mathcal{IV}_{4}$ denotes the set of irrelevant and invalid instruments. We propose a three step procedure to achieve the estimation of endogenous treatment effects. Firstly, we consider a nonparametric additive reduced form model and estimate it by adaptive group Lasso, which select relevant instruments (denoted by $\mathcal{A}_{R}^{*}$) consistently and estimate the optimal instruments $\mathbf{D}^{*}$. This data-driven approach can usually adopt the linearity form automatically for a truly linear reduced form model. In the second step, we replace the original treatment variable by its estimated value and select the valid instruments (denoted by $\mathcal{A}_{V}^{*}$) consistently by adaptive Elastic-Net. In the final step, we take the selected invalid instruments (denoted by $\mathcal{A}_{I}^{*}=(\mathcal{A}_{V}^{*})^{c}$) as covariates and run a least squares regression to obtain the treatment effect estimator. It is shown that our procedure can select relevant and invalid instruments consistently\footnote{In the common sense of IV selection, we need to select the strong and valid IV. In the perspective of variable selection using Lasso type procedures, as we will demonstrate in our model part, the invalid instruments have non-zero coefficients in the structural equation and they should be selected. When we say we can select the invalid IV, it is not to be confused with the ultimate goal of selecting the $\mathcal{IV}_{1}$.}. The estimator has the desired theoretical properties such as consistency and asymptotic normality. Therefore, it is called {\bf{R}}obust {\bf{IV}} {\bf{E}}stimator to both the {\bf{I}}nvalid and {\bf{I}}rrelevant instruments (R2IVE), where the number 2 in R2IVE denotes both types of bad instruments. Monte Carlo simulations demonstrate that our estimator performs better than the existing IV estimators (such as 2SLS, NAIVE \citep{Fan2018Nonparametric} and \emph{sisVIVE} \citep{Kang2015Instrumental}) for the endogenous treatment effects. In the empirical study, we revisit the classic question of trade and growth. It shows that our R2IVE estimators can select the invalid and relevant instruments and provide a more robust result than  \cite{Frankel1999Trade} and NAIVE \citep{Fan2018Nonparametric} when there exists potentially invalid instruments.

The paper is organized as follows. Section 2 introduces the model setting, identification and estimation. Section 3 presents theoretical results. Section 4 collects simulation results to evaluate the finite sample performance of the proposed estimator. In section 5, we illustrate the usefulness of our estimator by revisiting the trade and growth study. Section 6 concludes. Technical proofs are given in the Appendix.

The following notations are used throughout the paper. For any $n$ by $L$ matrix $\mathbf{X}$, we denote the $(i,j)$-th element of matrix $\mathbf{X}$ as $X_{ij}$, the $i$th row as $\mathbf{X}_{i.}$, and the $j$th column as $\mathbf{X}_{.j}$. $\mathbf{X}^{T}$ is the transpose of $\mathbf{X}$. $\mathbf{\mathcal{M}}_\mathbf{X} = \mathbf{I}_n-\mathbf{\mathcal{P}}_\mathbf{X}$, where $\mathbf{\mathcal{P}}_\mathbf{X} = \mathbf{X}(\mathbf{X}^{T}\mathbf{X})^{-1}\mathbf{X}^{T}$ is the projection matrix onto
the column space of $\mathbf{X}$, and $\mathbf{I}_n$ is the $n$-dimensional identity matrix. Let $\boldsymbol{\iota}_{s}$ denote a $1\times s$ vector of ones. The $l_p$-norm is denoted by $\left\|\cdot\right\|_p$, and the $l_0$-norm, $\left\|\cdot\right\|_0$, denotes the number
of non-zero components of a vector. $\left\|\cdot\right\|$ is $l_2$-norm following conventions. For any set $A$, we denote $A^{c}$ to be its complement and $|A|$ is the cardinality of set $A$. %We use $\left\|\cdot\right\|_{\infty}$ to denote the maximal element of a vector.

%%%%%%%%%%%%%%%%%%%%%%%%%%%%%%%%%%%%%%%%%%%%%%%%%%%%%%%%%%%%%%%%%%%%%%%%%%%%%%%%%%%%%

\section{Model}\label{model}
\subsection{Model Setting}
For $i=1,\ldots, n$, let $Y_i^{(0,\mathbf{0})}\in \mathbb{R}$ be the potential outcome without any treatment and instruments, $Y_{i}^{(d,\mathbf{z})} \in \mathbb{R}$ be the potential outcome if the individual $i$ were to have treatment $d$ and instruments values $\mathbf{z}$. $D_i^{(\mathbf{z})}\in \mathbb{R}$ is the endogenous treatment if the individual $i$ were to have instruments $\mathbf{z}$, $d,d^{\prime} \in \mathbb{R}$ are two possible values of the treatment, and $\mathbf{z},\mathbf{z}^{\prime} \in \mathbb{R}^{L}$ are two possible values of the instruments. We consider a true high-dimensional instrumental variable setting such that the dimensionality $L$ is allowed to be potentially larger than the sample size $n$. Without loss of generality, we consider the univariate endogenous treatment variable. Suppose we have the following potential outcomes model,
\begin{align}
Y_{i}^{\left(d^{\prime}, \mathbf{z}^{\prime}\right)}-Y_{i}^{(d, \mathbf{z})} &=\left(\mathbf{z}^{\prime}-\mathbf{z}\right)^{T} \boldsymbol{\alpha}_1^{*}+\left(d^{\prime}-d\right) \beta^{*} \label{eq1}\\
E\left(Y_{i}^{(0,\bf{0})} | \mathbf{Z}_{i .}\right) &=\mathbf{Z}_{i .}^{T} \boldsymbol{\alpha}_2^{*} \label{eq2}
\end{align}
where  $\beta^{*} \in \mathbb{R}$ is the treatment parameter of interest, $\boldsymbol{\alpha}_1^{*} \in \mathbb{R}^{L}$ represents the direct effect of $\mathbf{z}$ on $Y$, and $\boldsymbol{\alpha}_2^{*}\in \mathbb{R}^{L}$ represents the presence of unmeasured confounders that affect both the instruments and the outcomes. An instrument $Z_j$ therefore does not satisfy the exclusion condition if $\alpha_{1j} \neq 0$ and does not satisfy the exogeneity condition if $\alpha_{2j}\neq 0$.

For each individual, only one possible realization of $Y_i^{(d,\mathbf{z})}$ and $D_i^{(\mathbf{z})}$ are observed, which are denoted as $Y_i$ and $D_i$, respectively, with the observed instrument values $\mathbf{Z}_{i.}$. We study the endogenous treatment effect through a random sample $\left\{Y_i; D_i; \mathbf{Z}_{i.}^{T}\right\}_{i=1}^{n}$. Let $\boldsymbol{\alpha}^{*}=\boldsymbol{\alpha}_1^{*}+\boldsymbol{\alpha}_2^{*}$, and  $\varepsilon_{i}=Y_{i}^{(0,\bf{0})}-E\left[Y_{i}^{(0,\bf{0})} | \mathbf{Z}_{i .}\right]$. Combining (\ref{eq1}) and (\ref{eq2}), the baseline model is given by
\begin{equation}\label{eq3}
Y_{i}=\mathbf{Z}_{i.}^{T} \boldsymbol{\alpha}^{*}+D_{i} \beta^{*}+\varepsilon_{i} \quad E\left(\varepsilon_{i} | \mathbf{Z}_{i}\right)=0
\end{equation}
where $D_i$ is an endogenous treatment variable, that is, $E(\varepsilon_{i}|D_i)\neq 0$\footnote{Our study focuses on a known endogenous treatment effect model.  \cite{Guo2018b} study the testing endogeneity problem and propose a new test that  has better power than the Durbin-Wu-Hausman (DWH) test in high dimensions.}.
Note that the model can include exogenous variables $\mathbf{X}_{i} \in \mathbb{R}^{p}$ and if so we can replace the variables $Y_i$, $D_i$ and $\mathbf{Z}_{i.}$ with the residuals after regressing them on $\mathbf{X}$ (e.g., replace $\mathbf{Y}$ by $\mathcal{M}_{\mathbf{X}}\mathbf{Y}$) \citep{Zivit1998Inference}. For simplicity, we assume
that $\mathbf{Y}$, $\mathbf{D}$, and the columns of $\mathbf{Z}$ are centered, which can be obtained from a residual transformation with $\mathbf{X}$ containing only the constant term.

\textbf{Definition 1:} Instrument $j\in \left\{1,\ldots,L\right\}$ is valid if $\alpha_j^{*} =0$, which means the instrument $j$ satisfies both condition (A2) and (A3), and it is invalid if $\alpha_j^{*} \neq 0$. Let $\mathcal{A}_{V}^{*}$ denote the set of valid instruments and $\mathcal{A}_{I}^{*}=(\mathcal{A}_{V}^{*})^{c}$ denote  the set of invalid instruments.

We also extend the \emph{sisVIVE }\citep{Kang2015Instrumental} by considering a nonparametric additive reduced form model with a large number of possible instruments.
\begin{align}
  %D_{i} & =\mathbf{Z}_{i.}^{T}\boldsymbol{\gamma}^{*}+v_{i},\quad E\left(v_{i} | \mathbf{Z}_{i}\right)=0  \\
  D_{i} &=\sum_{j=1}^{L}f_{j}(Z_{ij})+\xi_{i},\quad E\left(\xi_{i} | \mathbf{Z}_{i.}\right)=0\label{eq4}
\end{align}
where $f_{j}(\cdot)$ is the $j$th unknown smooth univariate function and $\xi_{i}$'s are i.i.d random errors with mean
0 and finite variance. For the model identification, we assume that all functions $f_j(\cdot)$'s are centered, that is, $E[f_j(Z_j)] = 0$, $1 \leq j \leq p$, where $Z_j$ denotes the $j$th instrument. As it is more flexible and generally applicable than the ordinary linear model, the nonparametric additive reduced form model \eqref{eq4} could achieve a better approximation to the optimal instruments $D_{i}^{*}=E(D_i|\mathbf{Z}_i.)$ \citep{AmemiyaThe,NeweyEfficient}. The resulting  estimator based on \eqref{eq4} is expected to be more efficient compared to the linear IV estimator, which will be confirmed both theoretically and numerically in the later sections. %Note that, here we allow a nonlinear reduced form and many irrelevant instruments, which is the second point where we differ from \cite{Kang2015Instrumental}.

\textbf{Definition 2:} Instrument $j\in \left\{1,\ldots,L\right\}$ is a non-redundant IV that satisfies (A1),  if $f_j(Z_j)\neq 0$. Let $\mathcal{A}_{R}^{*}$ denote the set of these instruments that are able to approximate the conditional expectation of the endogenous
variable.

%\textbf{Definition 3:} Instrument $j\in \left\{1,\ldots,L\right\}$ is instruments if (A1)-(A3) are satisfied and we denote $\mathcal{IV}_{I}=\mathcal{IV}_{V}\bigcap\mathcal{IV}_{R}$ to be the set of instruments.
\subsection{Identification and Estimation}

%There are many methods to estimate the linear reduced form, such as Lasso \citep{Tibshirani1996Regression}, Adaptive Lasso \citep{Hui2006The}, square root Lasso \citep{Belloni2010}, SCAD \citep{Fan2001SCAD}, etc.
In this section, we illustrate the identification and estimation of models \eqref{eq3} and \eqref{eq4}. Without knowing which instrument is irrelevant or invalid, we consider a large set of candidate instruments, $\mathcal{IV}=\mathcal{IV}_{1}\cup\mathcal{IV}_{2}\cup\mathcal{IV}_{3}\cup\mathcal{IV}_{4}$ as introduced in Section \ref{intro}.  Here, $\mathcal{IV}_{1}$  is the set of ideal instruments satisfying the conditions (A1)-(A3). $\mathcal{IV}_{2}$ could contribute to the construction of optimal instruments and the correct specification of model \eqref{eq3}. $\mathcal{IV}_{3}$  should be excluded in the models, otherwise, the reduced form equation \eqref{eq4} will  provide poor finite sample properties of estimators and the structural model \eqref{eq3} will be overfitted. $\mathcal{IV}_{4}$ is the set of instruments that all conditions (A1)-(A3) are not satisfied. These instruments should be excluded from the model \eqref{eq4} but included in the model \eqref{eq3}. The model \eqref{eq3} will not be correctly specified if we delete these instruments mistakenly, that is, we take invalid instruments as valid. Our goal is therefore to select relevant instruments (denoted by $\mathcal{A}_{R}^{*}=\mathcal{IV}_{1}\cup\mathcal{IV}_{2}$) consistently for the model \eqref{eq4} and invalid instruments (denoted by $\mathcal{A}_{I}^{*}=\mathcal{IV}_{2}\cup\mathcal{IV}_{4}$) consistently for the model \eqref{eq3}. Specifically, we first estimate the equation \eqref{eq3} by adaptive group Lasso, and then substitute the estimated optimal instruments into \eqref{eq4} to select the valid instruments using adaptive Elastic-Net.

Many researchers have studied additive nonparametric models \citep{Linton1970}. The nonparametric study
of reduced form equation is often troubled by the curse of dimensionality \citep{NeweyEfficient}, which has been the focus of a substantial body of recent literature on high-dimensional problems. \cite{Huang2010} proposed a variable selection procedure in nonparametric additive models using the adaptive group Lasso based on a spline approximation to the nonparametric component. \cite{Fan2018Nonparametric} extend it to IV estimation, which selects the strong instruments consistently and adopts the linearity form automatically for a truly linear reduced form model. Here, we estimate the optimal instruments and select the relevant instruments consistently following \cite{Fan2018Nonparametric}.

Let $\mathcal{S}_n$ be the space of polynomial splines of degrees $h \geq 1$ and $\left\{\phi_{k},k=1,\ldots,m_n\right\}$ be
the normalized B-spline basis functions for $\mathcal{S}_n$, where $m_n$ is the
sum of the polynomial degree $h$ and the number of knots.
Let $\psi_k(Z_{ij}) = \phi_k(Z_{ij}) - n^{-1} \sum_{i=1}^{n}\phi_k(Z_{ij})$ be the centered B spline basis function for the $j$th instrument. Thus, for each $f_{nj} \in \mathcal{S}_{n}$, it can be represented by the linear combination of normalized B-spline series
\begin{equation}\label{eq5}
f_{nj}(Z_{ij})=\sum _{k=1}^{m_{n}}\gamma _{jk}\psi _{k}(Z_{ij}) \quad 1\leq j \leq L
\end{equation}
Under suitable smoothness conditions, the function $f_j(Z_{ij})$ in \eqref{eq4} can be well approximated by the function $f_{nj}(Z_{ij})$ in $\mathcal{S}_n$
by carefully choosing the coefficients $\left\{\gamma_{j1},...,\gamma_{jm_n}\right\}$ \citep{Stone1985}. Then the model (\ref{eq4}) can be rewritten as
\begin{equation}\label{eq6}
D_{i} \approx \sum_{j=1}^{L}\sum _{k=1}^{m_{n}}\gamma _{jk}\psi _{k}(Z_{ij})+\xi_{i}
\end{equation}

Denote $\mathbf{D}=\left(D_1,...,D_n\right)^{T}$ as the $n\times1$ vector of endogenous treatment variable. Let $\boldsymbol{\gamma}_j= (\gamma_{j1},\gamma_{j2},\ldots,\gamma_{jm_n})^{T}$ be the $m_n \times 1$  vector of parameters corresponding to the $j$th instrument in (\ref{eq6}) and $\boldsymbol{\gamma}=(\boldsymbol{\gamma}_1^{T},\ldots,\boldsymbol{\gamma}_L^{T})^{T}$ be the $m_nL\times1$ vector of parameters. Let $\boldsymbol{U}_{ij}=(\psi _{1}(Z_{ij}),...,\psi _{m_{n}}(Z_{ij}))^{T}$ be the $m_n\times1$ vector, $\boldsymbol{U}_{j}=(\boldsymbol{U}_{1j},...,\boldsymbol{U}_{nj})^{T}$ be the $n\times m_n$ design matrix for the $j$th instrument and $ \boldsymbol{U}=(\boldsymbol{U}_{1},...,\boldsymbol{U}_{L})$ be the corresponding $n\times m_{n}L$ design matrix. To select the significant instruments and estimate the component functions simultaneously, we consider the following penalized objective function with an adaptive group Lasso penalty
\begin{equation}\label{eq7}
\widehat{\boldsymbol{\gamma}}_{n}=\arg\min \limits_{\boldsymbol{\gamma}}\left \{ \left \| \mathbf{D}-\boldsymbol{U}\boldsymbol{\gamma } \right \|_{2}^{2} +\lambda _{n}\sum _{j=1}^{L}\omega_{j}\left \| \boldsymbol{\gamma }_{j} \right \|_{2}\right \}
\end{equation}
where the weights is defined by
\begin{equation}\label{eq8}
\omega_{j}=\left\{\begin{matrix}
                \left \| \boldsymbol{\widetilde{\gamma} }_{j} \right \|_{2}^{-1}& \textup{if}\left \| \boldsymbol{\widetilde{\gamma} }_{j} \right \|_{2}> 0\\
                \infty &  \textup{if}\left \| \boldsymbol{\widetilde{\gamma} }_{j} \right \|_{2}= 0
                \end{matrix}\right.
\end{equation}
and  $\widetilde{\boldsymbol{\gamma}}_{n}=(\widetilde{\boldsymbol{\gamma}}_{1}^{T},\ldots,
\widetilde{\boldsymbol{\gamma}}_{L}^{T})^{T}$ is obtained from  group Lasso
\begin{equation}\label{eq9}
\widetilde{\boldsymbol{\gamma}}_{n}=\arg\min \limits_{\boldsymbol{\gamma}}\left \{ \left \| \mathbf{D}-\boldsymbol{U}\boldsymbol{\gamma } \right \|_{2}^{2} +\lambda _{n0}\sum _{j=1}^{L}\left \| \boldsymbol{\gamma }_{j} \right \|_{2}\right \}
\end{equation}
Denote the $\widehat{\mathcal{A}}_{R}=\left\{j:\left\|\widehat{\boldsymbol{\gamma}}_j\right\|^{2}>0\right\}$, and the adaptive group Lasso estimators of  $f_j$ in \eqref{eq4} are
\begin{equation*}
\widehat{f}_{nj}(Z_{ij})=\sum_{k=1}^{m_n}\widehat{\gamma}_{jk}\psi_{k} (Z_{ij}),\quad j \in \widehat{\mathcal{A}}_{R}
\end{equation*}
Therefore the endogenous variable $D_i$ can be estimated by
\begin{equation}\label{eq10}
\widehat{D}_{i}=\sum _{j\in \widehat{\mathcal{A}}_R}\sum _{k=1}^{m_{n}}\widehat{\gamma} _{jk}\psi _{k}(Z_{ij})
\end{equation}
Denote $\widehat{\mathbf{D}}=(\widehat{D}_{1},\cdots,\widehat{D}_{n})^{T}$, then $\widehat{\mathbf{D}}$ is the optimal instrument similar to \cite{Belloni2012sparse} and \cite{Fan2018Nonparametric} when $\left|\mathcal{A}_{R}^{*}\right|$ is greater than 1, which is shown in the following Theorem \ref{thm}. Note that  when the nonparametric
additive reduced form model \eqref{eq4} is indeed a linear model, this data-adaptive approach
can usually select $m_n = 1$ to degenerate to a linear model of \eqref{eq4}. The EBIC \citep{Chen2008Extended} and BIC \citep{Hansheng2007Tuning} could be used to choose the tuning parameters $\lambda_{0n}$, $\lambda_{n}$, and $m_n$ adaptively in practice for the high-dimensional and low-dimensional cases, respectively.

Next, we illustrate how to select the valid instruments and estimate the true endogenous treatment effect $\beta^{*}$. By taking
the conditional expectation of both sides of \eqref{eq3} given instrumental variables $\mathbf{Z}_i$, we have
\begin{equation}\label{eq11}
E(Y_{i}|\mathbf{Z}_{i.})=D_{i}^{*} \beta^{*}+\mathbf{Z}_{i.}^{T} \boldsymbol{\alpha}^{*}
\end{equation}
where $D_i^{*}=E(D_i|\mathbf{Z_{i.}})$. Denote $\nu_i = Y_i - E(Y_{i}|\mathbf{Z}_{i.})$, it is straightforward to show that $E(\nu_{i})=E[E(\nu_{i}|\mathbf{Z}_{i.})]=0$ and cov$(D_i^{*}\nu_{i})=E[D_i^{*}E(\nu_{i}|\mathbf{Z}_{i.})]=0$. Adding $\nu_i$ to both side of \eqref{eq11}, we have
\begin{equation}\label{eq12}
Y_{i}=D_{i}^{*} \beta^{*}+\mathbf{Z}_{i.}^{T} \boldsymbol{\alpha}^{*}+\nu_{i}
\end{equation}
Thus, $D_i^{*}$ is an exogenous variable in \eqref{eq12}. It is worth noting that
the coefficient of the optimal instrument $D_i^{*}$ in the model \eqref{eq12} remains the same $\beta^{*}$ as  in the structural equation \eqref{eq3}. If $D_i^{*}$ is known, the model \eqref{eq12} can be easily
estimated by linear estimation method. In practice, we replace $D_i^{*}$ by its estimate $\widehat{D}_i$ to obtain the final IV estimator for $\beta^{*}$. Substituting $\widehat{D}_{i}$ from \eqref{eq10} into \eqref{eq12}, we have
\begin{equation}\label{eq13}
\mathbf{Y}=\widehat{\mathbf{D}} \beta^{*} +\mathbf{Z} \boldsymbol{\alpha}^{*}+\widehat{\boldsymbol{\nu}}
\end{equation}
Here, we need to have at least one relevant and valid instrument so that the equation \eqref{eq13} is not troubled by the problem of potential collinearity\footnote{This is the assumption needed for the inference of the endogenous treatment effect in our model. There is a rich literature on the inference using many weak IVs or invalid IVs. In recent studies, \cite{Hansen2014} consider IV estimation with many weak IVs in high-dimensional models where the consistent model selection in the first stage may not be possible. \cite{Bietal2020} discuss inferring treatment effects after testing instrument strength in linear models. \cite{Berkowitz2012} shows how valid inferences can be made when an instrumental variable does not perfectly satisfy the orthogonality condition. }. We consider a  two step procedure to estimate the equation \eqref{eq13}. Notice that unlike $\beta^{*}$, we are essentially concerned with which element in $\boldsymbol{\alpha}^{*}$ is not equal to 0 (invalid IV) but to a lesser extent, the true value of $\boldsymbol{\alpha}^{*}$.
\begin{enumerate}
  \item [(S1)] We first remove the effect of $\widehat{\mathbf{D}}$ from \eqref{eq13}. Multiplying by $\mathcal{M}_{\widehat{\mathbf{D}}}$ on both sides of \eqref{eq13}, we have
\begin{equation}\label{eq14}
\mathcal{M}_{\widehat{\mathbf{D}}}\mathbf{Y}=\mathcal{M}_{\widehat{\mathbf{D}}}\mathbf{Z} \boldsymbol{\alpha}^{*} +\mathcal{M}_{\widehat{\mathbf{D}}}\widehat{\boldsymbol{\nu}}
\end{equation}
Denote $\widetilde{\mathbf{Y}}= \mathcal{M}_{\widehat{\mathbf{D}}}\mathbf{Y}$, $\widetilde{\mathbf{Z}}= \mathcal{M}_{\widehat{\mathbf{D}}}\mathbf{Z}$ and $\widetilde{\boldsymbol{\nu}}= \mathcal{M}_{\widehat{\mathbf{D}}}\widehat{\boldsymbol{\nu}}$. The equation \eqref{eq14} can be written as $\widetilde{\mathbf{Y}}=\widetilde{\mathbf{Z}}\boldsymbol{\alpha}^{*} +\widetilde{\boldsymbol{\nu}}$, which is a standard high-dimensional linear model. The popular linear high-dimensional methods include the Lasso \citep{Tibshirani1996Regression}, SCAD \citep{Fan2001SCAD}, Elastic-Net \citep{Elasticnet2005}, group Lasso \citep{groupLasso2006}, adaptive Lasso \citep{Hui2006The}, Dantzig selector
\citep{Candes2007} and adaptive Elastic-Net \citep{Zou2009Adenet} among others. Here, we use the adaptive Elastic-Net proposed in \cite{Zou2009Adenet} to estimate the $\boldsymbol{\alpha}^{*}$ consistently.
\begin{enumerate}
  \item We first compute the Elastic-Net estimator $\widetilde{\alpha}$ \citep{Elasticnet2005} by \eqref{eq15}, and then construct the adaptive weights by \eqref{eq16}.
\begin{equation}\label{eq15}
\widetilde{\boldsymbol{\alpha}} = (1+\frac{\lambda_{2}}{n})\left\{\arg\min \limits_{\boldsymbol{\alpha}}\left\|\widetilde{\mathbf{Y}}- \widetilde{\mathbf{Z}}\boldsymbol{\alpha} \right\|_{2}^{2}+\lambda_2\left\|\boldsymbol{\alpha}\right\|_{2}^{2}+ \lambda_1 \left\|\boldsymbol{\alpha}\right\|_1\right\}
\end{equation}
\begin{equation}\label{eq16}
\widehat{\omega}_{j}=\left\{\begin{matrix}
                \left|\widetilde{\alpha}_{j}\right|^{-\tau}& \textup{if } \widetilde{\alpha}_{j}> 0\\
                \infty &  \textup{if } \widetilde{\alpha}_{j}= 0
                \end{matrix}\right.
\end{equation}
where we take $\tau=\lceil\frac{2\eta}{1-\eta}\rceil$ following \cite{Zou2009Adenet}\footnote{Here the constant $0\le \eta<1$. See Assumption 2 (A3) in the appendix.}.
  \item Then we solve the following optimization problem to get the adaptive Elastic-Net estimates
\begin{equation}\label{eq17}
\widehat{\boldsymbol{\alpha}} = (1+\frac{\lambda_{2}}{n})\left\{\arg\min \limits_{\boldsymbol{\alpha}}\left\|\widetilde{\mathbf{Y}}- \widetilde{\mathbf{Z}}\boldsymbol{\alpha} \right\|_{2}^{2}+\lambda_2\left\|\boldsymbol{\alpha}\right\|_{2}^{2}+ \lambda_1^{*} \widehat{\omega}_{j}\left\|\boldsymbol{\alpha}\right\|_1\right\}
\end{equation}
Denote $\widehat{\mathcal{A}}_{I}= \left\{Z_j, \text{ for } j: \left|\alpha_{j}\right| > 0 \right\}$ as the empirical set of invalid IVs. Note that the $\ell_1$ regularization parameters, $\lambda_1$ and $\lambda_1^{*}$ are allowed to be different while the $\ell_2$ regularization parameters are same. Here, we first use a proper range of values with (relatively small) grid for $\lambda_2$. Then, for each $\lambda_2$, the EBIC and BIC are used to choose the $\lambda_1$ and $\lambda_1^{*}$.  This method produces the  entire solution path of the adaptive Elastic-Net.
\end{enumerate}

  \item [(S2)]  In step 2 we take the selected invalid instruments as covariates in \eqref{eq13} and run a least square regression. The resulting IV estimator of $\beta^{*}$ takes the form
\begin{equation}\label{estimator}
\widehat{\beta} =\left(\widehat{\mathbf{D}}^{T}\mathcal{M}_{\widehat{\mathcal{A}}_{I}}\widehat{\mathbf{D}}\right)^{-1}\widehat{\mathbf{D}}^{T}\mathcal{M}_{\widehat{\mathcal{A}}_{I}}\mathbf{Y}
\end{equation}
\end{enumerate}

In summary, we present the following Algorithm 1 for the estimator.
\begin{algorithm*}
\caption{\textbf{R}obust \textbf{IV} \textbf{E}stimator to both the \textbf{I}nvalid and \textbf{I}rrelevant instruments (R2IVE)}
\begin{algorithmic}
\vspace{0.1cm}
\STATE {\bf Step 1.} Obtain the penalized estimator $\widehat{\boldsymbol{\gamma}}_{n}$ in \eqref{eq7} and estimate the conditional expectation of the endogenous treatment $\widehat{D}_{i}$ in \eqref{eq10}, where the weights is defined by \eqref{eq8} and \eqref{eq9} and the BIC or EBIC are applied to choose the tuning parameters $\lambda_n$, $\lambda_{n0}$ and $m_n$. \vspace{0.2cm}
\STATE {\bf Step 2.} Take $\widetilde{\mathbf{Y}}= \mathcal{M}_{\widehat{\mathbf{D}}}\mathbf{Y}$, $\widetilde{\mathbf{Z}}= \mathcal{M}_{\widehat{\mathbf{D}}}\mathbf{Z}$ and $\widetilde{\boldsymbol{\nu}}= \mathcal{M}_{\widehat{\mathbf{D}}}\widehat{\boldsymbol{\nu}}$. Obtain the penalized estimator $\widehat{\boldsymbol{\alpha}}$ and the invalid instruments set $\widehat{\mathcal{A}}_{I}$ in \eqref{eq17}, where the weights is defined by \eqref{eq15} and \eqref{eq16}. For a proper range of values for $\lambda_2$, the BIC or EBIC are used to choose the $\lambda_1$ and $\lambda_1^{*}$ for each $\lambda_2$.\vspace{0.2cm}
\STATE {\bf Step 3.} Take the selected  invalid instruments as covariates and run a least square regression for \eqref{eq13}. The resulting IV estimator of $\beta^{*}$ takes form as $\widehat{\beta} =\left(\widehat{\mathbf{D}}^{T}\mathcal{M}_{\widehat{\mathcal{A}}_{I}}\widehat{\mathbf{D}}\right)^{-1}\widehat{\mathbf{D}}^{T}\mathcal{M}_{\widehat{\mathcal{A}}_{I}}\mathbf{Y}$.
\vspace{0.1cm}
\end{algorithmic}
\end{algorithm*}

It is important to consider the model selection in the asymptotic theory of the estimator as discussed in \cite{chernozhukov2018}.

\section{Theoretical Properties}
We assume the following regularity conditions for theoretical study. %Firstly, we need to clarify the regularity condition for the nonparametric estimation.
\begin{assumption}
(C1) $\sqrt{E(D_i^{2})} < \infty$ for $i=1,...,n$.

(C2) The true value $\beta^{*}$ is bounded, $|\beta^{*}| \leq C$. The number of the relevant instruments $s_1=\left|\mathcal{A}_{R}^{*}\right|$ is fixed and $s_{1}$ is a positive integer. The number of invalid instruments $s_{2} = \left|\mathcal{A}_{I}^{*}\right|$ is less than $L/2$. The number of relevant and valid instruments $\left|\mathcal{IV}_{I}\right| \ge 1$ .

(C3) The distribution of $\xi_i$ has subexponential tails, $E[exp(C|\xi_i)] < \infty$ for a finite positive constant $C$.  $E(\varepsilon_{i}^{3})$ is bounded away from zero and the infinity. $E[|\nu_i|^{2+\delta}]<\infty$ for some $\delta>0$.

%The function of $\mathbf{D}^{*}$ admits an approximately sparse form, namely $\boldsymbol{\gamma}^{*}$ such that $D_{i}^{*}=\mathbf{U}_{i}'\boldsymbol{\gamma}^{*}+r_{i}$, $ \left\|\boldsymbol{\gamma}^{*}\right\|_{0} \leq s_{1}'$, where $s_{1}'$ is a positive integer greater than 1.

%(C4) $\sqrt{E(\nu_i^{2})} < \infty$ for $i=1,...,n$.

%(C4) Define the minimal and maximal m-sparse eigenvalues of a semi-definite matrix $\mathbf{M}$ as $\phi_{\min}(m)[\mathbf{M}] := \min\limits_{1 \leq \left\|\boldsymbol{\delta}\right\|_{0}\leq m }\frac{\boldsymbol{\delta}'\mathbf{M}\boldsymbol{\delta} }{\left\|\boldsymbol{\delta}\right\|^{2}}$ and $\phi_{\max}(m)[\mathbf{M}] := \max\limits_{1 \leq \left\|\boldsymbol{\delta}\right\|_{0}\leq m }\frac{\boldsymbol{\delta}'\mathbf{M}\boldsymbol{\delta} }{\left\|\boldsymbol{\delta}\right\|^{2}}$. There is an absolute sequence $a_n\rightarrow \infty$ such that with a high probability the maximal and minimal sparse eigenvalues are bounded from above and away from zero. Namely with probability at least $1-\Delta_{n}$,
%\begin{equation}\label{assm1}
 % b_2 \leq \phi_{\min}(a_ns'_1)\left(\mathbf{U}'\mathbf{U}/n\right) \leq \phi_{\max}(a_ns'_1)\left(\mathbf{U}'\mathbf{U}/n\right)\leq B_2
%\end{equation}
%\begin{equation}\label{assm1}
 % b_2 \leq \phi_{\min}(a_ns_2)\left(\mathbf{Z}'\mathbf{Z}/n\right) \leq \phi_{\max}(a_ns_2)\left(\mathbf{Z}'\mathbf{Z}/n\right)\leq B_2
%\end{equation}
%where $0<b_2<B_2<\infty$ are absolute constants.
\end{assumption}

Here we only give the assumptions used in the proof of Theorem \ref{thm}. The assumptions  of Lemma \ref{lem3.1} and \ref{lem3.2} are put in the appendix. The condition (C1) imposes mild restriction on the finite second moment of the endogenous treatment variable $D_i$. Condition (C2) restricts the boundedness of true treatment effect and the number of relevant instruments and invalid instruments. The number of invalid instruments is needed to be less than the half of instruments, which is the identification condition in \cite{Kang2015Instrumental}. Condition (C3) clarifies the conditions about error terms. The distribution of the random errors $\xi_i$'s should not be too heavy-tailed, and it is satisfied for $\xi_i$'s that are bounded uniformly or normally distributed. %clarifies the sparsity form of $\mathbf{D}^{*}$ and restricts the moments of approximation errors. %Condition (C4) has been proposed as Condition SE(P) for Sparse Eigenvalues in \cite{belloni2014inference}. This condition can directly holds for i.i.d. zero-mean sub-Gaussian random vectors or i.i.d. bounded zero-mean random vectors \citep{rudelson2013reconstruction}. This condition is more general than a standard assumption that the population Gram matrix has eigenvalues bounded from above and away from zero.
\begin{lem}\label{lem3.1}
Using the group Lasso estimator $\widetilde{\gamma}$ with $\lambda_{n0}\asymp O(\sqrt{n\log(m_nL)})$ and $m_n\asymp O(n^{1/(2s+1)})$ to construct the weight
for the adaptive group Lasso estimator\footnote{Here $s$ is a positive constant. See Assumption 2 (A2) in the appendix.}. Suppose Assumption 1 and 2 hold and $\lambda_{n}\asymp O(\sqrt{n})$, then
\begin{equation}\label{lem1}
P\left(\widehat{\mathcal{A}}_{R}=\mathcal{A}_{R}\right) \rightarrow 1, \text{as } n \rightarrow \infty
\end{equation}
\begin{equation}\label{lem2}
\sum_{j \in \mathcal{A}_{R}}\left\|\widehat{\boldsymbol{\gamma}_{nj}}-\boldsymbol{\gamma}_{j }\right\|_{2}^{2}=O_{p}\left(n^{-(2 s-1) /(2 s+1)}\right)
\end{equation}
\begin{equation}\label{lem3}
\sum_{j \in \mathcal{A}_{R}}\left\|\widehat{f}_{n j}-f_{j}\right\|_{2}^{2}=O_{p}\left(n^{-2 s /(2 s+1)}\right)
\end{equation}
\begin{equation}\label{lem4}
\left\|\mathbf{D}^{*}-\hat{\mathbf{D}}\right\|=o_p(1)
\end{equation}
\end{lem}

This lemma shows the selection and estimation consistency of the adaptive group Lasso for high-dimensional nonparametric additive reduced form model, which essentially follows the results of Theorem 3 in \cite{Huang2010}. We give the proof of equation \eqref{lem4} in the appendix.
\begin{lem}\label{lem3.2}
Under the Assumption 1 and 2, the adaptive Elastic-Net achieves the model selection consistency, that is, $P\left(\widehat{\mathcal{A}}_{I}=\mathcal{A}_{I}\right) \rightarrow 1, \text{as } n \rightarrow \infty$.
\end{lem}

This lemma shows the selection consistency of the adaptive Elastic-Net for high-dimensional linear structure model. That is, the true set of invalid instruments can be identified with probability tending to 1. The proof of this lemma is in the appendix.
\begin{thm}\label{thm}
Suppose Assumption 1 and 2 hold, the R2IVE estimator in \eqref{estimator} is root-$n$ consistent and asymptotically normal. That is
\begin{equation}\label{thm2}
 \sigma_{n}^{-1}\sqrt{n}\left(\widehat{\beta}-\beta^{*}\right)\rightarrow N(0,1)
\end{equation}
where the asymptotic variance $\sigma_{n}^{2}$ are given in the following two cases.
\begin{itemize}
  \item[(i)] In the case that the structural error is heteroscedastic, there exists constants $k_1$ such that $E(\nu_{i}^{2}|\mathbf{Z_{i.}})\leq k_1$ holds almost surely for $1\leq i \leq n$,
      $$\sigma_{n}^{2}=\left[E\left(\mathbf{D}^{*T} \mathcal{M}_{\mathcal{A}_{I}}\mathbf{D}^{*}\right)\right]^{-1}E\left[\mathbf{D}^{*T} \mathcal{M}_{\mathcal{A}_{I}}\mathbf{D}^{*}\nu_{i}^{2}\right]\left[E\left(\mathbf{D}^{*T} \mathcal{M}_{\mathcal{A}_{I}}\mathbf{D}^{*}\right)\right]^{-1}$$
  \item[(ii)] In the case that the structural error is homoscedastic, that is, $E(\nu_i^2|\mathbf{Z_{i.}}) = \sigma_{\nu}^{2}$ almost surely for all $1 \leq i \leq n$, \eqref{estimator} holds with $\sigma_{n}^{2}=\sigma_{\nu}^{2}\left[E\left(\mathbf{D}^{*T} \mathcal{M}_{\mathcal{A}_{I}}\mathbf{D}^{*}\right)\right]^{-1}.$
\end{itemize}
\end{thm}

\section{Simulation}
  We conduct various simulation studies to evaluate the estimation performance for different methods. We consider a structural model with one endogenous variable,
  \begin{equation*}
  Y_{i}=D_{i} \beta^{*}+\mathbf{Z}_{i.}^{T} \boldsymbol{\alpha}^{*}+\varepsilon_{i}
  \end{equation*}
  where $\beta^{*}=0.75$, $\mathbf{Z}_{i .}=(Z_{i1},Z_{i2},\ldots,Z_{iL})^{T}$ is generated from a multivariate normal distribution $N(0,\mathbf{\Sigma)}$, and $\mathbf{\Sigma}=\left(\rho_{j_{1} j_{2}}\right)_{L \times L}$ with $\rho_{j_{1} j_{2}}=0.5^{\left|j_{1}-j_{2}\right|}$, for $j_{1}, j_{2}=1, \ldots, L,$ and for each $i=1, \ldots, n$. We set $\boldsymbol{\alpha}=(\mathbf{0}_{q},\boldsymbol{\iota}_{s_2},\mathbf{0}_{L-q-s_2})^{T}$, where $\boldsymbol{\iota}_{s_2}$ is a $1 \times s_2$ vector of ones, which means the first $q$ and the last $L-q-s_2$ instruments are valid, and $s_2$ is the number of invalid instruments.
  The endogenous variable is generated based on either of the following two reduced form models,
  \begin{align*}
  \text{Model 1:}\quad D_{i}&=\mathbf{Z}_{i.}\boldsymbol{\gamma}^{*}+\xi_{i}\\
  \text{Model 2:} \quad D_{i}&=2Z_{i1}^{2}+0.75Z_{i2}^{2}+1.5Z_{i3}^{2}+3\sin(\pi Z_{i4})+\xi_{i}
  \end{align*}
For the model 1, we consider a ``cut-off at $s_1$" design, that is,  $\boldsymbol{\gamma}^{*}=(2, 0.75, 1.5, 1, \dots , \boldsymbol{0}_{L-s_1})^{T}$, where $\mathbf{0}_{L-s_1}$ is a $1\times (L-s_1)$ vector of zeros and $s_1$ is the number of strong instruments. We fill in the values of non-zero elements in $\boldsymbol{\gamma}^{*}$ by replicating the non-zero elements of $(2,0.75,1.5,1)$ to the until its length is $s_1$. E.g, if $s_1=6$, the non-zero elements of $\boldsymbol{\gamma}^{*}$ are $(2,0.75,1.5,1, 2,0.75)$. For the model 2, the main setting is that the number of non-zero nonlinear components is fixed at 4 as shown above. And to evaluate the performance when the number of strong IVs increase, we also replicate the same functional forms for $Z_{i5},\dots , Z_{i12}$ in the form of $Z_{i1},\dots ,Z_{i4}$ in one simulation setting. We generate the error terms in both the structural model and reduced form models by
  $$\left(\begin{array}{c}{\varepsilon_{i}} \\ {\xi_{i}}\end{array}\right) \stackrel{\mathrm{i.i.d.}}{\sim} N\left(\left[\begin{array}{c}{0} \\ {0}\end{array}\right],\left[\begin{array}{cc}{1} & {0.8} \\ {0.8} & {1}\end{array}\right]\right)$$

In the simulations, we vary (i) the sample size $n$, (ii) the number of strong instruments $s_1$, (iii) the number of invalid instruments $s_2$ and (iv) the distribution among the four instruments sets $\mathcal{IV}_{1}-\mathcal{IV}_{1}$, which is  affected by the value of $q$. The model settings are summarized in Table \ref{tab5}.

\begin{table}[htbp!]
\centering
\caption{The Model Setting}
\label{tab5}
\begin{tabular}{ccccccccccc}
\hline

  & $n$ & $L$ & $s_1$ & $s_2$ & $q$ & $|\mathcal{IV}_1|$ & $|\mathcal{IV}_2|$ & $|\mathcal{IV}_3|$ & $|\mathcal{IV}_4|$ \\ \hline
\multirow{6}{*}{Linear} & 200 & 100 & 10 & 0 & 10 & 10 & 0 & 90 & 0 \\ %\hline
&200 & 100 & 10 & 10 & 7 & 7 & 3 & 83 & 7 \\ %\hline
&200 & 100 & 10 & 30 & 7 & 7 & 3 & 63 & 27 \\ %\hline
%&200 & 100 & 10 & 60 & 7 & 7 & 3 & 33 & 57 \\ %\hline
&200 & 100 & 4 & 30 & 2 & 2 & 2 & 68 & 28 \\ %\hline
&200 & 100 & 20 & 30 & 14 & 14 & 6 & 56 & 24 \\ %\hline
&200/500/1000 & 100 & 20 & 20 & 14 & 14 & 6 & 66 & 14 \\ \hline
\multirow{3}{*}{Nonlinear}&500/200 & 100 & 4 & 0 & 4 & 4 & 0 & 96 & 0 \\ %\hline
&500 & 100 & 4 & 20 & 2 & 2 & 2 & 78 & 18 \\ %\hline
&500 & 100 & 12 & 20 & 9 & 9 & 3 & 71 & 17 \\ \hline

\end{tabular}
\end{table}

  We run each simulation $R = 1000$ times and compute the average of the estimation bias (denoted by "Bias"), $R^{-1} \sum_{r=1}^{R}\left(\widehat{\beta}^{r}-\beta^{*}\right)$,  with its empirical standard deviation and the estimated mean squared errors (denoted by "MSE"), $R^{-1} \sum_{r=1}^{R}\left(\widehat{\beta}^{r}-\beta^{*}\right)^{2}$, where $\widehat{\beta}^{r}$ denotes an estimator of ${\beta}^{*}$ in the $r$th experiment. We compare our method with OLS, 2SLS, Oracle 2SLS (this is the 2SLS using the oracle $\mathcal{IV}_1$ set), NAIVE \citep{Fan2018Nonparametric} and \emph{sisVIVE} \citep{Kang2015Instrumental}. For 2SLS and NAIVE, we report the estimation results using only the endogenous treatment variable in the structural equation. We also present results for the so-called Post \emph{sisVIVE} estimator (sisVIVE.post in the following tables), which is 2SLS estimator that takes the set of nonzero estimated coefficients $\boldsymbol{\alpha}$ in \emph{sisVIVE} as covariates in the structural equation. We report the model selection performance for the estimators that embedding variable selection. Specifically, we report average number ("mean") of instruments selected as strong (for NAIVE) or invalid (for \emph{sisVIVE}, sisVIVE.post, and R2IVE) together with the minimum, median and  maximum numbers of invalid IV selection, and the proportion of  the instruments selected as strong or invalid to all strong or invalid instruments ("freq"), respectively. Notice all the model settings in Table \ref{tab5} are the "Stronger Valid" cases when the valid instruments are stronger than the invalid instruments. The readers could refer to \cite{Kang2015Instrumental} for more discussion on this setting.

  All simulation studies are conducted using the statistical software R. In particular, the R package \emph{naivereg} \citep{Fan2018Nonparametric} is used to estimate the optimal instruments and the R package \emph{sisVIVE} \citep{Kang2015Instrumental} is used to get the \textit{sisVIVE} estimator. We use the R package \emph{gcdnet} \citep{Yang2012An} to select invalid instruments with (adaptive) Elastic-Net penalty. The EBIC \citep{Chen2008Extended} and BIC \citep{Hansheng2007Tuning} methods are employed to find the optimal tuning parameters for the high-dimensional and low-dimensional cases, respectively. 

\subsection{Linear reduced form equation}

In this setting, we consider the linear reduced form. We first fix the sample size $n$, IV dimension $L$, the number of relevant instruments $s_1$ and the value of $q$ but change  the number of invalid instruments $s_2$ to check the influence of the number of invalid instruments. Then we fix $n$,  $L$ and  $s_2$ but change the value of $s_1$ and $q$ to check the influence of the strength of instruments. Finally, we fix $L$, $s_1$ and $s_2$ but change the sample size. 

\begin{table}[htbp!]
\centering
\caption{Fix $n=200,L=100,s_1=10$, change the number of invalid instruments $s_2$}
\label{tab1}
\begin{tabular}{cccccccccc}
\hline
                          &              & Bias    & std dev & MSE    & mean                   & median              & max                 & min                 & freq                  \\ \hline
\multirow{7}{*}{$s_2=0$}  & OLS          & 0.0163  & 0.0102  & 0.0004 &                        &                     &                     &                     &                       \\
                          & 2SLS         & 0.0079  & 0.0102  & 0.0002 &                        &                     &                     &                     &                       \\
                          & Oracle 2SLS  & 0.0003  & 0.0103  & 0.0001 &                        &                     &                     &                     &                       \\
                          & \emph{NAIVE }       & 0.0038  & 0.0105  & 0.0001 & 12.08                  & 12                  & 16                  & 10                  & 1                  \\
                          & \emph{sisVIVE}      & 0.0080   &     0.0103    & 0.0002 & \multirow{2}{*}{0.08}  & \multirow{2}{*}{0}  & \multirow{2}{*}{19} & \multirow{2}{*}{0}  & \multirow{2}{*}{-} \\
                          & sisVIVE.post & 0.0081  & 0.0102  & 0.0002 &                        &                     &                     &                     &                       \\
                          & \textbf{R2IVE}        & \textbf{0.0005}  & \textbf{0.0103}  & \textbf{0.0001} & \textbf{0.37}                   & \textbf{0}                   & \textbf{3}                   & \textbf{0}                   & -                \\ \hline
\multirow{7}{*}{$s_2=10$} & OLS          & 0.5310   & 0.0983  & 0.2927 &                        &                     &                     &                     &                       \\
                          & 2SLS         & 0.5282  & 0.0988  & 0.2899 &                        &                     &                     &                     &                       \\
                          & Oracle 2SLS  & 0.0008  & 0.0138  & 0.0002 &                        &                     &                     &                     &                       \\
                          & \emph{NAIVE }       & 0.5217  & 0.1010   & 0.2838 & 12.17                  & 12                  & 16                  & 10                  & 1                  \\
                          & \emph{sisVIVE}      & 0.2755  &    0.1290     & 0.0926 & \multirow{2}{*}{47.17} & \multirow{2}{*}{49} & \multirow{2}{*}{78} & \multirow{2}{*}{10} & \multirow{2}{*}{1} \\
                          & sisVIVE.post & 0.4846  & 0.0834  & 0.2766 &                        &                     &                     &                     &                       \\
                          & \textbf{R2IVE }        & \textbf{0.0019}  & \textbf{0.0137}  & \textbf{0.0002} & \textbf{10.53}                  & \textbf{10 }                 & \textbf{14}                  & \textbf{10}                  & \textbf{1}                 \\ \hline
\multirow{7}{*}{$s_2=30$} & OLS          & 0.2686  & 0.0937  & 0.0806 &                        &                     &                     &                     &                       \\
                          & 2SLS         & 0.2629  & 0.0942  & 0.0778 &                        &                     &                     &                     &                       \\
                          & Oracle 2SLS  & 0.0002  & 0.0145  & 0.0002 &                        &                     &                     &                     &                       \\
                          & \emph{NAIVE }       & 0.2541  & 0.0963  & 0.0737 & 12.14                  & 12                  & 17                  & 10                  & 1                 \\
                          & \emph{sisVIVE }     & 0.1796  &    0.1153     & 0.0456 & \multirow{2}{*}{59.45} & \multirow{2}{*}{63} & \multirow{2}{*}{85} & \multirow{2}{*}{32} & \multirow{2}{*}{0.997}  \\
                          & sisVIVE.post & 0.3122  & 0.0798  & 0.1440  &                        &                     &                     &                     &                       \\
                          & \textbf{R2IVE}         & \textbf{-0.0005} & \textbf{0.015}   & \textbf{0.0003} & \textbf{33.63 }                 & \textbf{33}                  & \textbf{51}                  & \textbf{30}                  & \textbf{1}                  \\ \hline
%\multirow{7}{*}{$s_2=60$} & OLS          & 0.2708  & 0.1356  & 0.0915 &                        &                     &                     &                     &                       \\
                          %& 2SLS         & 0.2651  & 0.1363  & 0.0888 &                        &                     &                     &                     &                       \\
                          %& Oracle 2SLS  & 0.0009  & 0.0161  & 0.0003 &                        &                     &                     &                     &                       \\
                          %& \emph{NAIVE }       & 0.2518  & 0.1394  & 0.083  & 12.11                  & 12                  & 16                  & 10                  & 0.999                   \\
                          %& \emph{sisVIVE}      & 0.0795  &    -     & 0.0093 & \multirow{2}{*}{75.6}  & \multirow{2}{*}{75} & \multirow{2}{*}{99} & \multirow{2}{*}{63} & \multirow{2}{*}{1} \\
                          %& sisVIVE.post & 0.1039  & 0.0487  & 0.0188 &                        &                     &                     &                     &                       \\
                          %& \textbf{R2IVE}         & \textbf{-0.0078} & \textbf{\textcolor[rgb]{1.00,0.00,0.00}{2.0313} } & \textbf{0.002}  & \textbf{66.36}                  & \textbf{66}                  & \textbf{84}                  & \textbf{59}                  & \textbf{0.993}                   \\ \hline
\end{tabular}
\begin{tablenotes}
\item[1] NOTE: This table summarizes the averages of estimated bias with the standard deviations and MSE, the average number of instruments selected as  strong (for NAIVE) or invalid (for \emph{sisVIVE}, sisVIVE.post and R2IVE)  together with the  median, minimum and  maximum numbers, and the proportion of  the instruments selected as strong or invalid to all strong or invalid instruments for respective estimators.   Notice \emph{sisVIVE} and sisVIVE.post share the same invalid IV selection results. The R2IVE share the same strong IV selection performance as the NAIVE.
\end{tablenotes}
\end{table}

\begin{figure}[htbp!]
  \centering
  \begin{minipage}{7cm}
     \includegraphics[width=7cm]{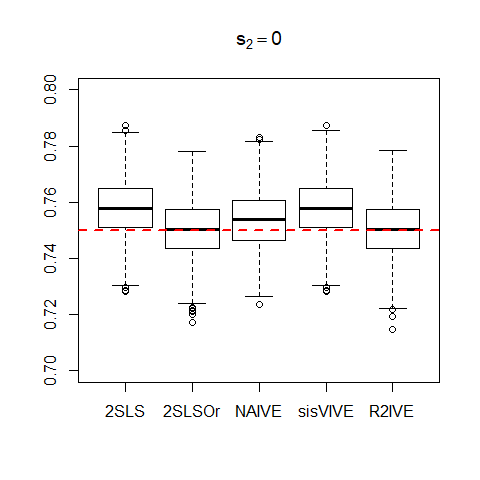}
  \end{minipage}
  %\begin{minipage}{7cm}
     %\includegraphics[width=7cm]{200_100_10_10.png}
  %\end{minipage}
  \begin{minipage}{7cm}
     \includegraphics[width=7cm]{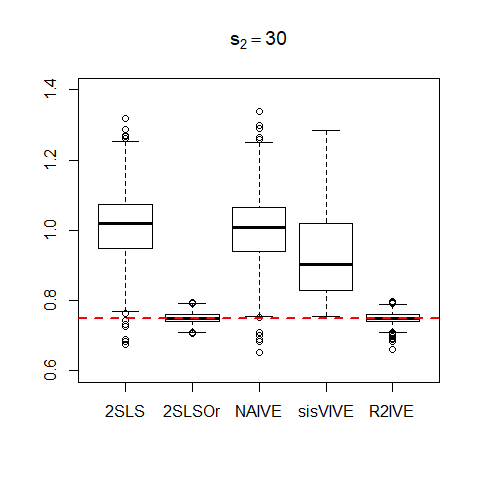}
  \end{minipage}
  %\begin{minipage}{7cm}
     %\includegraphics[width=7cm]{200_100_10_60.png}
  %\end{minipage}
  \caption{Fix $n=200,L=100,s_1=10$, change the number of invalid instruments $s_2$}
  \label{fig1}
\end{figure}

In the first case, we fix $n=200$ and  $L=100$. The number of relevant instruments $s_1=10$. We use the values of $s_2=0, 10, 30$ to check the influence of the number of invalid instruments on estimation results. For $s_2=0$, we have $|\mathcal{IV}_{2}|=|\mathcal{IV}_{4}|=0$. Set $|\mathcal{IV}_{1}|=10$ and $|\mathcal{IV}_{3}|=90$.  For other nonzero values of $s_2$, we set $q=7$, which means $|\mathcal{IV}_{1}|=7$, $|\mathcal{IV}_{2}|=3$, $|\mathcal{IV}_{3}|=100-s_2-7$ and $|\mathcal{IV}_{4}|=s_2-3$. The specific numbers of each type of instrument variables are summarized in Table \ref{tab5}. Table \ref{tab1} reports the estimation bias with the standard deviations and MSE, as well as the model selection results. Figure \ref{fig1} shows the box plots of bias for these estimators. In Figure \ref{fig1}, we do not include the results of OLS estimators since they are always biased and have the largest MSE, which will enlarge the scale of the figure and make the differences between other IV estimators less distinguishable. When there is no invalid instruments, the \emph{sisVIVE} is outperformed by the NAIVE due to the effect of many irrelevant instruments. When the invalid instruments exist, the 2SLS and NAIVE estimators perform similarly and are both severely biased due to the effects of invalid instruments. The \emph{sisVIVE} estimators have smaller bias and MSE compared to 2SLS and NAIVE estimators. However, they are still substantially biased although the invalid instruments are always selected as invalid by \emph{sisVIVE} and the sisVIVE.post does not help reducing bias. Furthermore, \emph{sisVIVE} always select too many invalid instruments, which reduce the efficiency of estimators. Our R2IVE performs best (among the non-Oracle estimators) with the smallest bias and MSE. It is very close to oracle 2SLS estimator in linear reduced form models and is shown to be robust to both irrelevant and invalid instruments. %Note that both \emph{sisVIVE} and our estimator become unstable when $s_2=60$ ($s_2>L/2$), where exist many outliers and the MSE values increase.
\begin{table}[htbp!]
\centering
\caption{Fix $n=200,L=100,s_2=30$, change the number of irrelevant instruments $s_1$}
\label{tab2}
\begin{tabular}{cccccccccc}
\hline
                          &              & Bias    & std dev & MSE    & mean                   & median              & max                 & min                 & freq                  \\ \hline
\multirow{7}{*}{$s_1=4$}  & OLS          & 0.5575  & 0.1670  & 0.3391 &                        &                     &                     &                     &                       \\
                          & 2SLS         & 0.5490  & 0.1699  & 0.3317 &                        &                     &                     &                     &                       \\
                          & Oracle 2SLS  & -0.0014 & 0.0336  & 0.0011 &                        &                     &                     &                     &                       \\
                          & NAIVE        & 0.5318  & 0.1764  & 0.3149 & 5.47                   & 5                   & 8                   & 4                   & 1                  \\
                          & \emph{sisVIVE }     & 0.6869  &    0.0421     & 0.4736 & \multirow{2}{*}{38.52} & \multirow{2}{*}{38} & \multirow{2}{*}{57} & \multirow{2}{*}{31} & \multirow{2}{*}{0.82}   \\
                          & sisVIVE.post & 0.6919  & 0.0382  & 0.4806 &                        &                     &                     &                     &                       \\
                          & \textbf{R2IVE}         & \textbf{0.0133}  & \textbf{0.0995}  & \textbf{0.0165} & \textbf{37.17}                  & \textbf{36}                  & \textbf{70 }                 & \textbf{30}                  & \textbf{1}                  \\ \hline
\multirow{7}{*}{$s_1=10$} & OLS          & 0.2686  & 0.0937  & 0.0806 &                        &                     &                     &                     &                       \\
                          & 2SLS         & 0.2629  & 0.0942  & 0.0778 &                        &                     &                     &                     &                       \\
                          & Oracle 2SLS  & 0.0002  & 0.0145  & 0.0002 &                        &                     &                     &                     &                       \\
                          & \emph{NAIVE }       & 0.2541  & 0.0963  & 0.0737 & 12.14                  & 12                  & 17                  & 10                  & 1                 \\
                          & \emph{sisVIVE}      & 0.1796  &    0.1155     & 0.0456 & \multirow{2}{*}{59.45} & \multirow{2}{*}{63} & \multirow{2}{*}{85} & \multirow{2}{*}{32} & \multirow{2}{*}{0.997}  \\
                          & sisVIVE.post & 0.3122  & 0.0798  & 0.1440 &                        &                     &                     &                     &                       \\
                          & \textbf{R2IVE}         & \textbf{-0.0005} & \textbf{0.0150}  & \textbf{0.0003} & \textbf{33.63}                  & \textbf{33 }                 & \textbf{51}                  & \textbf{30}                  & \textbf{1}                 \\ \hline
\multirow{7}{*}{$s_1=20$} & OLS          & 0.2441  & 0.0640  & 0.0636 &                        &                     &                     &                     &                       \\
                          & 2SLS         & 0.2413  & 0.0642  & 0.0622 &                        &                     &                     &                     &                       \\
                          & Oracle 2SLS  & 0.0008  & 0.0094  & 0.0001 &                        &                     &                     &                     &                       \\
                          & NAIVE        & 0.2379  & 0.0653  & 0.0608 & 21.70                  & 21                  & 30                  & 20                  & 1                 \\
                          & \emph{sisVIVE}      & 0.0340  &    0.0156     & 0.0014 & \multirow{2}{*}{40.73} & \multirow{2}{*}{40} & \multirow{2}{*}{80} & \multirow{2}{*}{30} & \multirow{2}{*}{1} \\
                          & sisVIVE.post & 0.0365  & 0.0142  & 0.0020 &                        &                     &                     &                     &                       \\
                          & \textbf{R2IVE}         & \textbf{0.0011}  & \textbf{0.0095}  & \textbf{0.0001} & \textbf{32.06}                  & \textbf{32}                  & \textbf{38}                  & \textbf{29}                  & \textbf{0.998}                   \\ \hline

\end{tabular}
\begin{tablenotes}
\item[1] Please see table notes in Table \ref{tab1}.
\end{tablenotes}
\end{table}
\begin{figure}[hbtp!]
  \centering
  \begin{minipage}{7cm}
     \includegraphics[width=7cm]{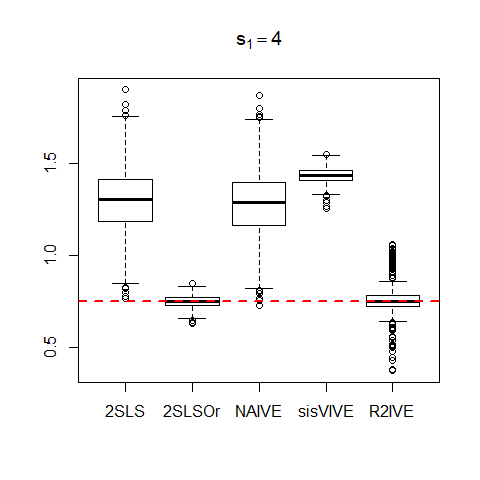}
  \end{minipage}
  \begin{minipage}{7cm}
     \includegraphics[width=7cm]{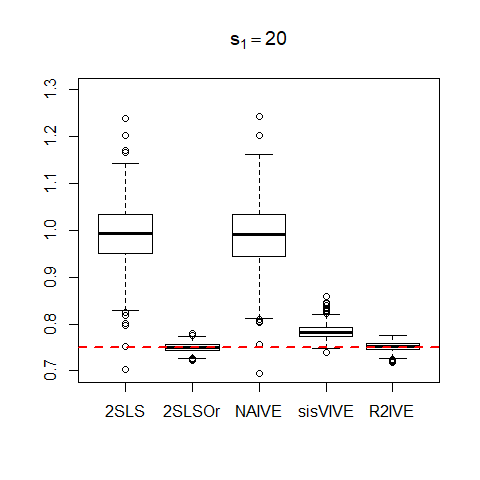}
  \end{minipage}
  \caption{Fix $n=200,L=100,s_2=30$, change the number of irrelevant instruments $s_1$}
  \label{fig2}
\end{figure}

In the second case, we fix $n=200$, $L=100$ and $s_2=30$ but change the value of $s_1=4, 10, 20$ to check the influence of the strength of instruments on different estimators. We set $q=2,7,14$ for different $s_1$, respectively.  $|\mathcal{IV}_{1}|=2,7,14$, $|\mathcal{IV}_{2}|=2,3,6$, $|\mathcal{IV}_{3}|=68,80,56$ and $|\mathcal{IV}_{4}|=28,27,24$ for different $s_1$ respectively. The results are shown in the Table \ref{tab2} and Figure \ref{fig2}. We observe that the \emph{sisVIVE} estimators are dependent of the number of relevant instruments, and they have diminishing bias and MSE when the number of relevant instruments $s_1$ increases. When $s_1=4$, the \emph{sisVIVE} is even outperformed by NAIVE that takes all instruments as valid. The sisVIVE.post does not reduce the bias of the \emph{sisVIVE}. This shows the importance of selecting strong IVs. Our R2IVE  performs best and can estimate the casual effect precisely. The R2IVE also improves as the number of relevant instruments increases.
% Please add the following required packages to your document preamble:
% \usepackage{multirow}
\begin{table}[!htbp]
\centering
\caption{Fix $L=100, s_1=20, s_2=20$, change the number size $n$}
\label{tab3}
\begin{tabular}{cccccccccc}
\hline
                          &              & Bias    & std dev & MSE    & mean                   & median              & max                 & min                 & freq                  \\ \hline
%\multirow{7}{*}{$n=100$}  & OLS          & 0.2399  & 0.0730  & 0.0624 &                        &                     &                     &                     &                       \\
                         % & 2SLS         & 0.2398  & 0.0730  & 0.0624 &                        &                     &                     &                     &                       \\
                          %& Oracle 2SLS  & 0.0013  & 0.0139  & 0.0002 &                        &                     &                     &                     &                       \\
                          %& \emph{NAIVE }       & 0.2058  & 0.0934  & 0.0577 & 21.67                  & 21                  & 29                  & 20                  & 1                 \\
                          %& \emph{sisVIVE}      & 0.0732  &    -     & 0.0091 & \multirow{2}{*}{27.93} & \multirow{2}{*}{27} & \multirow{2}{*}{70} & \multirow{2}{*}{20} & \multirow{2}{*}{1} \\
                          %& sisVIVE.post & 0.0856  & 0.0215  & 0.0165 &                        &                     &                     &                     &                       \\
                          %& \textbf{R2IVE}         & \textbf{0.0212}  & \textbf{0.0509}  & \textbf{0.0083} & \textbf{21.21}                  & \textbf{21}                  & \textbf{27}                  & \textbf{20}                  & \textbf{1}                  \\ \hline
\multirow{7}{*}{$n=200$}  & OLS          & 0.2450  & 0.0508  & 0.0626 &                        &                     &                     &                     &                       \\
                          & 2SLS         & 0.2422  & 0.0509  & 0.0613 &                        &                     &                     &                     &                       \\
                          & Oracle 2SLS  & 0.0007  & 0.0092  & 0.0001 &                        &                     &                     &                     &                       \\
                          & \emph{NAIVE }       & 0.2395  & 0.0518  & 0.0601 & 21.67                  & 21                  & 29                  & 20                  & 1                 \\
                          & \emph{sisVIVE}      & 0.0334  &    0.0136     & 0.0013 & \multirow{2}{*}{27.93} & \multirow{2}{*}{27} & \multirow{2}{*}{70} & \multirow{2}{*}{20} & \multirow{2}{*}{1} \\
                          & sisVIVE.post & 0.0315  & 0.0126  & 0.0015 &                        &                     &                     &                     &                       \\
                          & \textbf{R2IVE}         & \textbf{0.0011}  & \textbf{0.0091}  & \textbf{0.0001} & \textbf{21.21}                  & \textbf{21}                  & \textbf{27}                  & \textbf{20}                  & \textbf{1}                  \\ \hline
\multirow{7}{*}{$n=500$}  & OLS          & 0.2418  & 0.0322  & 0.0595 &                        &                     &                     &                     &                       \\
                          & 2SLS         & 0.2373  & 0.0324  & 0.0573 &                        &                     &                     &                     &                       \\
                          & Oracle 2SLS  & 0.0002  & 0.0056  & 0.0000 &                        &                     &                     &                     &                       \\
                          & NAIVE        & 0.2346  & 0.0328  & 0.0561 & 25.79                  & 26                  & 31                  & 21                  & 0.999                   \\
                          & \emph{sisVIVE}      & 0.0176  &    0.0095     & 0.0004 & \multirow{2}{*}{24.09} & \multirow{2}{*}{23} & \multirow{2}{*}{50} & \multirow{2}{*}{20} & \multirow{2}{*}{1} \\
                          & sisVIVE.post & 0.0117  & 0.0069  & 0.0002 &                        &                     &                     &                     &                       \\
                          & \textbf{R2IVE}         & \textbf{0.0005}  & \textbf{0.0056}  & \textbf{0.0000} & \textbf{20.56}                  & \textbf{20}                  & \textbf{24}                  & \textbf{20}                  & \textbf{1}                  \\ \hline
\multirow{7}{*}{$n=1000$} & OLS          & 0.2424  & 0.0227  & 0.0593 &                        &                     &                     &                     &                       \\
                          & 2SLS         & 0.2373  & 0.0228  & 0.0568 &                        &                     &                     &                     &                       \\
                          & Oracle 2SLS  & -0.0002 & 0.0039  & 0.0000 &                        &                     &                     &                     &                       \\
                          & NAIVE        & 0.2352  & 0.0231  & 0.0559 & 25.72                  & 26                  & 31                  & 21                  & 1                \\
                          & \emph{sisVIVE}      & 0.0123  &   0.0070      & 0.0002 & \multirow{2}{*}{25.56} & \multirow{2}{*}{22} & \multirow{2}{*}{99} & \multirow{2}{*}{20} & \multirow{2}{*}{1} \\
                          & sisVIVE.post & 0.0064  & 0.0056  & 0.0001 &                        &                     &                     &                     &                       \\
                          & \textbf{R2IVE}         & \textbf{-0.0001} & \textbf{0.0039}  & \textbf{0.0000} & \textbf{20.40}                  & \textbf{20}                  & \textbf{24}                  & \textbf{20}                  & \textbf{1}                 \\ \hline
%\multirow{7}{*}{$n=2000$} & OLS          & 0.2421  & 0.0162  & 0.0589 &                        &                     &                     &                     &                       \\
                          %& 2SLS         & 0.2367  & 0.0163  & 0.0563 &                        &                     &                     &                     &                       \\
                          %& Oracle 2SLS  & 0.0000  & 0.0028  & 0.0000 &                        &                     &                     &                     &                       \\
                          %& NAIVE        & 0.2348  & 0.0164  & 0.0554 & 25.58                  & 25                  & 32                  & 21                  & 1                 \\
                          %& \emph{sisVIVE}      & 0.0091  &    -     & 0.0002 & \multirow{2}{*}{29.37} & \multirow{2}{*}{22} & \multirow{2}{*}{99} & \multirow{2}{*}{20} & \multirow{2}{*}{0.995}  \\
                          %& sisVIVE.post & 0.0046  & 0.0087  & 0.0001 &                        &                     &                     &                     &                       \\
                          %& \textbf{R2IVE}         & \textbf{0.0000}  & \textbf{0.0028}  & \textbf{0.0000} & \textbf{20.29}                  & \textbf{20 }                 & \textbf{23}                  & \textbf{20 }                 & \textbf{1}                 \\ \hline
\end{tabular}
\begin{tablenotes}
\item[1] Please see table notes in Table \ref{tab1}.
\end{tablenotes}
\end{table}
\begin{figure}[htbp!]
  \centering
  %\begin{minipage}{7cm}
  %   \includegraphics[width=7cm]{n=100.png}
  %\end{minipage}
  \begin{minipage}{7cm}
     \includegraphics[width=7cm]{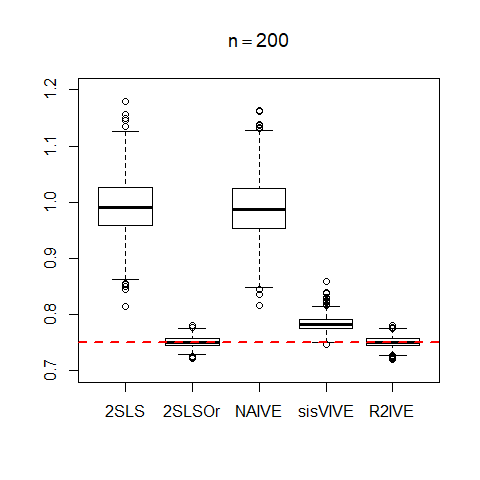}
  \end{minipage}
  %\begin{minipage}{7cm}
     %\includegraphics[width=7cm]{n=500.png}
  %\end{minipage}
  \begin{minipage}{7cm}
     \includegraphics[width=7cm]{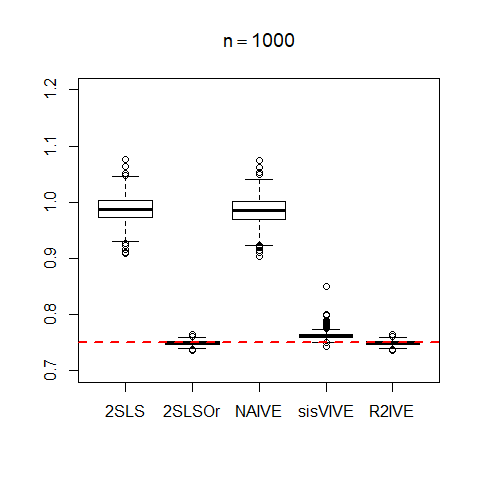}
  \end{minipage}
  \caption{Fix $L=100, s_1=20, s_2=20$, change the sample size $n$}
  \label{fig3}
\end{figure}

In the last case, we fix $L=100$,  $s_1=20$,  $s_2=20$ and $q=14$  while changing the sample size. The results are shown in the Table \ref{tab3} and Figure \ref{fig3}. The increase of the sample size does not improve the estimated performance of 2SLS and NAIVE, as they are always biased due to the endogeneity of IVs. The \emph{sisVIVE} estimators have diminishing bias and MSE when the sample size increases. Our R2IVE estimator always performs best with the smallest bias and MSE. Compared to \emph{sisVIVE}, our method has much better finite sample performance. It is very close to oracle 2SLS in linear models. Note that the sisVIVE.post can reduce some bias from \emph{sisVIVE} when $n$ is 1000 but is still outperformed by our estimator.

\subsection{Nonlinear reduced form equation}

\begin{figure}[]
  \centering
  \begin{minipage}{7cm}
     \includegraphics[width=7cm]{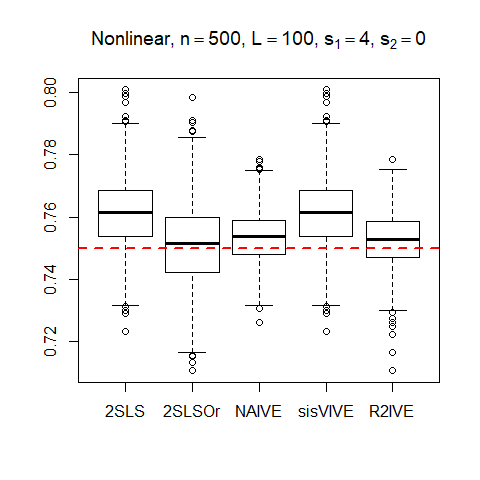}
  \end{minipage}
  \begin{minipage}{7cm}
     \includegraphics[width=7cm]{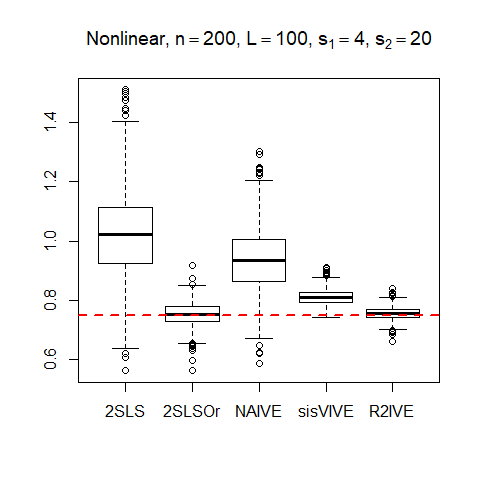}
  \end{minipage}

  \begin{minipage}{7cm}
     \includegraphics[width=7cm]{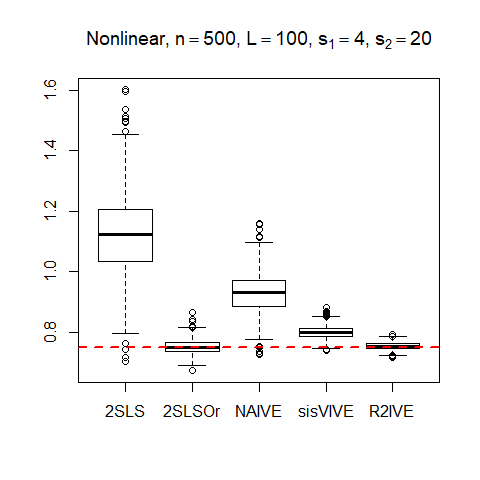}
  \end{minipage}
  \begin{minipage}{7cm}
     \includegraphics[width=7cm]{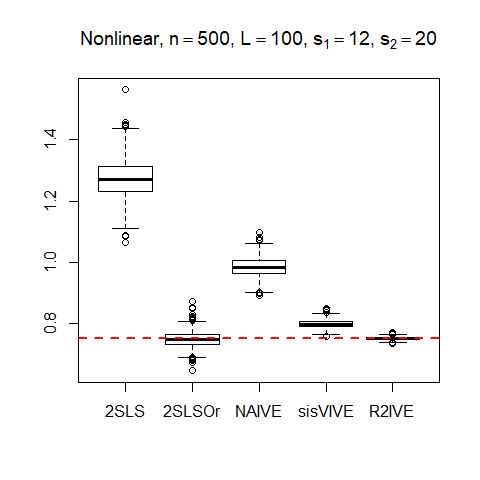}
  \end{minipage}
  \caption{Nonlinear reduced form setting}\label{fig4}
\end{figure}

In this subsection, we consider the nonlinear reduced form. The results are summarized in Table \ref{tab4} and Figure \ref{fig4}. The influence of the invalid instruments is checked by comparing the top left and bottom left plots of Figure \ref{fig4}. Similar to the linear case, the 2SLS and NAIVE estimators become biased due to the invalid instruments. When there is no invalid instruments, the \emph{sisVIVE} is outperformed by NAIVE since it does not consider the irrelevant instruments and  the nonlinear reduced form equation. When the invalid instruments exist, our R2IVE always performs best. Different from the linear case, the NAIVE  always outperforms the 2SLS by capturing the nonlinear structure. We check the influence of the strength of instruments by comparing the bottom left and bottom right plots of Figure \ref{fig4}. Both \emph{sisVIVE} and R2IVE are improved when the number of relevant instruments increase and our estimators always outperform \emph{sisVIVE} and sisVIVE.post. We check the influence of sample size by comparing the top right with bottom left plot of Figure \ref{fig4}. Both \emph{sisVIVE} and R2IVE improve as the sample size increases in the "stronger invalid" settings. 
\begin{table}[]
\centering
\caption{Nonlinear reduced form equation}
\label{tab4}
\begin{tabular}{cccccccccc}
\hline
         &              & Bias    & std dev & MSE    & mean                   & median              & max                 & min                 & freq               \\ \hline
         & OLS          & 0.0256  & 0.0079  & 0.0007 &                        &                     &                     &                     &                    \\
         & 2SLS         & 0.0111  & 0.0115  & 0.0003 &                        &                     &                     &                     &                    \\
$n=500$,   & Oracle 2SLS  & 0.0011  & 0.0136  & 0.0002 &                        &                     &                     &                     &                    \\
$p=100$,   & NAIVE        & 0.0035  & 0.0081  & 0.0001 & 5.66                   & 6                   & 9                   & 4                   & 1                  \\
$s_1=4$,  & \emph{sisVIVE}      & 0.0111  & 0.0113  & 0.0003 & \multirow{2}{*}{0.01}  & \multirow{2}{*}{0}  & \multirow{2}{*}{8}  & \multirow{2}{*}{0}  & \multirow{2}{*}{-}                 \\
$s_2=0$   & sisVIVE.post & 0.0112  & 0.0116  & 0.0003 &                        &                     &                     &                     &                    \\
         & \textbf{R2IVE}        & \textbf{0.0025}  & \textbf{0.0140}  & \textbf{0.0001} & \textbf{0.40}                   & \textbf{0 }                  & \textbf{5}                   & \textbf{0}                   & -                  \\ \hline
         & OLS          & 0.1999  & 0.0952  & 0.0500 &                        &                     &                     &                     &                    \\
         & 2SLS         & 0.2771  & 0.1166  & 0.0989 &                        &                     &                     &                     &                    \\
$n=200$,   & Oracle 2SLS  & 0.0013  & 0.0402  & 0.0015 &                        &                     &                     &                     &                    \\
$p=100$,   & NAIVE        & 0.1895  & 0.0987  & 0.0471 & 5.70                   & 6                   & 8                   & 3                   & 0.901              \\
$s_1=4$,  & \emph{sisVIVE}      & 0.0602  & 0.0259  & 0.0043 & \multirow{2}{*}{25.03} & \multirow{2}{*}{24} & \multirow{2}{*}{44} & \multirow{2}{*}{20} & \multirow{2}{*}{1} \\
$s_2=20$  & sisVIVE.post & 0.0345  & 0.0207  & 0.0018 &                        &                     &                     &                     &                    \\
         & \textbf{R2IVE}        & \textbf{0.0052}  & \textbf{0.0305}  & \textbf{0.0005} & \textbf{23.82}                  & \textbf{23}                  & \textbf{48}                  & \textbf{20}                  & \textbf{1}                  \\ \hline
         & OLS          & 0.1972  & 0.0597  & 0.0428 &                        &                     &                     &                     &                    \\
         & 2SLS         & 0.3715  & 0.0875  & 0.1550 &                        &                     &                     &                     &                    \\
$n=500$,   & Oracle 2SLS  & 0.0009  & 0.0237  & 0.0006 &                        &                     &                     &                     &                    \\
$p=100$,   & NAIVE        & 0.1799  & 0.0608  & 0.0364 & 5.67                   & 6                   & 9                   & 4                   & 1                  \\
$s_1=4$,  & \emph{sisVIVE}      & 0.0486  & 0.0211  & 0.0028 & \multirow{2}{*}{22.34} & \multirow{2}{*}{22} & \multirow{2}{*}{35} & \multirow{2}{*}{20} & \multirow{2}{*}{1} \\
$s_2=20$  & sisVIVE.post & 0.0287  & 0.0184  & 0.0014 &                        &                     &                     &                     &                    \\
         & \textbf{R2IVE}        & \textbf{0.0037}  & \textbf{0.0169}  & \textbf{0.0002} & \textbf{22.05}                  & \textbf{22}                  & \textbf{31}                  & \textbf{20}                  & \textbf{1}                  \\ \hline
         & OLS          & 0.2373  & 0.0288  & 0.0572 &                        &                     &                     &                     &                    \\
         & 2SLS         & 0.5223  & 0.0475  & 0.2770 &                        &                     &                     &                     &                    \\
$n=500$,   & Oracle 2SLS  & -0.0008 & 0.0248  & 0.0006 &                        &                     &                     &                     &                    \\
$p=100$,   & NAIVE        & 0.2346  & 0.0290  & 0.0560 & 17.72                  & 18                  & 22                  & 12                  & 1                  \\
$s_1=12$, & \emph{sisVIVE}      & 0.0487  & 0.0136  & 0.0026 & \multirow{2}{*}{31.12} & \multirow{2}{*}{30} & \multirow{2}{*}{63} & \multirow{2}{*}{21} & \multirow{2}{*}{1} \\
$s_2=20$  & sisVIVE.post & 0.0219  & 0.0124  & 0.0007 &                        &                     &                     &                     &                    \\
         & \textbf{R2IVE}        & \textbf{0.0020}  & \textbf{0.0084}  & \textbf{0.0000} & \textbf{20.28}                  & \textbf{20}                  & \textbf{24}                  & \textbf{20}                  & \textbf{1}                  \\ \hline
\end{tabular}%
\begin{tablenotes}
\item[1] NOTE: In the last section of this table, we replicate the same functional forms for $Z_{i5},\dots , Z_{i12}$ in the form of $Z_{i1},\dots ,Z_{i4}$ so that the number of strong IVs $s_1=12$. Other table notes please refer to those in Table \ref{tab1}.
\end{tablenotes}
\end{table}

%\begin{figure}
  %\centering
  %\begin{minipage}{7cm}
     %\includegraphics[width=7cm]{linear_low_dimension.png}
  %\end{minipage}
  %\begin{minipage}{7cm}
     %\includegraphics[width=7cm]{linear_high_dimension.png}
  %\end{minipage}

  %\begin{minipage}{7cm}
     %\includegraphics[width=7cm]{non_low_dimension.png}
  %\end{minipage}
  %\begin{minipage}{7cm}
     %\includegraphics[width=7cm]{non_high_dimension.png}
  %\end{minipage}
  %\caption{The comparison between high and low dimensions}
  %\label{fig5}
%\end{figure}
\section{Applications to Trade and Economic Growth}
\subsection{Background}
In this section, we illustrate the usefulness of our estimator by revisiting the classic question of trade and growth. The effect of trade on growth is a very important research topic in both theoretical and empirical economics, which has strong effect on trade policies.

One important issue in the empirical study of trade and growth is the endogeneity of trade variable due to the unobserved common driving forces that cause both trade and growth. \citet[\emph{FR99} henceforth]{Frankel1999Trade} circumvented the endogeneity problem utilizing instrumental variable constructed using gravity model of trade \citep{Anderson1979}. They showed that trade activities positively correlate with growth rate using cross-sectional data from 150 countries and economies using data from the mid-1980s. In \emph{FR99}, they consider a linear structural equation, which include the log of GDP per worker (outcome variable), the share of international trade to GDP (explanatory variable of interests) and two exogenous variables representing the size of a country: population and land area. They also consider a linear bilateral trade reduced form equation, where the instruments include the distance between two countries, dummy variables for landlocked countries, common border between two countries and the interaction terms, and two exogenous variable aforementioned. The instrumental variable (called proxy for trade in \emph{FR99}) is the sum of predicted bilateral trade shares for country $i$.
\cite{Fan2018Nonparametric} extend the study of \emph{FR99} by considering more potential instruments and considering a nonlinear reduced form. Besides the instrument used in \emph{FR99}, they also include total water area, coastline, the arable land as percentage of total land, land boundaries, forest area as percentage of land area, the number of official and other commonly used languages in a country, and the interaction terms of constructed trade proxy with these variables (in total 15 instruments). The selected instruments include the proxy for trade (the original instrument in \emph{FR99}), area of land, total population, and the interaction term of proxy for trade and number of languages. The NAIVE method provides stronger results regarding trade on growth than \emph{FR99}. In this study, we are concerned about the invalid IVs, which means some instruments might affect growth directly. The inclusion of invalid instruments leads to the inconsistency of $\beta$.
\subsection{Data and Model Setting}
Following \emph{FR99} and NAIVE, we use the cross-sectional data from 158 countries and economies and update the data to year 2017 to investigate the contemporary effect of trade on growth. We consider a linear structural equation
\begin{equation}\label{emp1}
ln Y_{i}= c + \beta D_{i}+\boldsymbol{\alpha}\mathbf{Z}_{i}+\boldsymbol{\delta}\mathbf{S}_i +\varepsilon_{i}
\end{equation}
where $Y_i$ is GDP per worker in country $i$, $D_i$ is the share of international trade to GDP, $\mathbf{S}_i$ is the size of country: population and Land area (same as \emph{FR99}), $\mathbf{Z_i}$ is the instruments. Besides the instruments used in \cite{Fan2018Nonparametric}, we also also include a instrumental variable related to air pollution: the density of PM 2.5.  \cite{Kukla-GryzEconomic} found that international trade and per capita income lead to the increase in air pollution in developing countries. In order to reduce the negative impact of international trade on the environment, the state will gradually adopt new policies with more environmental friendly standards hence raising the costs of production, which means air pollution could in turn affect international trade. On the other hand, there is empirical evidence for the environmental Kuznets curve between economic growth and environmental pollution. \cite{Ali2018Pollution} conclude the impact of  pollution abatement on economic growth could turn into win-win policy options. Hence, there is some reasons to believe that PM2.5 may impact trade, but also affect economic growth through mechanisms other than trade. $\varepsilon_{i}$ is unobserved random disturbances in the growth function. 

The reduced form model we consider is
\begin{equation}\label{emp2}
D_{i}=\mu+\sum_{j}f_{j}(z_{ij})+\xi_{i}
\end{equation}
where $f_{j}(\cdot)$ is the $j$th unknown smooth univariate functions, $z_{ij}$ is the $i$th observed value of the aforementioned $j$th instrument and $\xi_{i}$ is unobserved random disturbances, which is likely to be correlated with $\varepsilon_{i}$.

Note that we can replace the variables $Y_i$, $D_i$, and $\mathbf{Z}_{i.}$ with the residuals after regressing them on $\mathbf{S}_i$ (e.g., replace $\mathbf{Y}$ by $\widetilde{\mathbf{Y}}=(\mathbf{I}-\mathcal{P}_{\mathbf{S}})\mathbf{Y}$). The equation \eqref{emp1} and \eqref{emp2} becomes
\begin{equation}\label{emp3}
\begin{aligned}
ln \widetilde{Y}_{i}&= c + \beta \widetilde{D}_{i}+\boldsymbol{\alpha}\widetilde{\mathbf{Z}}_{i} +\varepsilon_{i}\\
\widetilde{D}_{i} &= \mu + \sum_{j}f_{j}(\widetilde{z}_{ij})+\xi_{i}
\end{aligned}
\end{equation}

\begin{figure}
  \centering
  % Requires \usepackage{graphicx}
  \includegraphics[width=10cm]{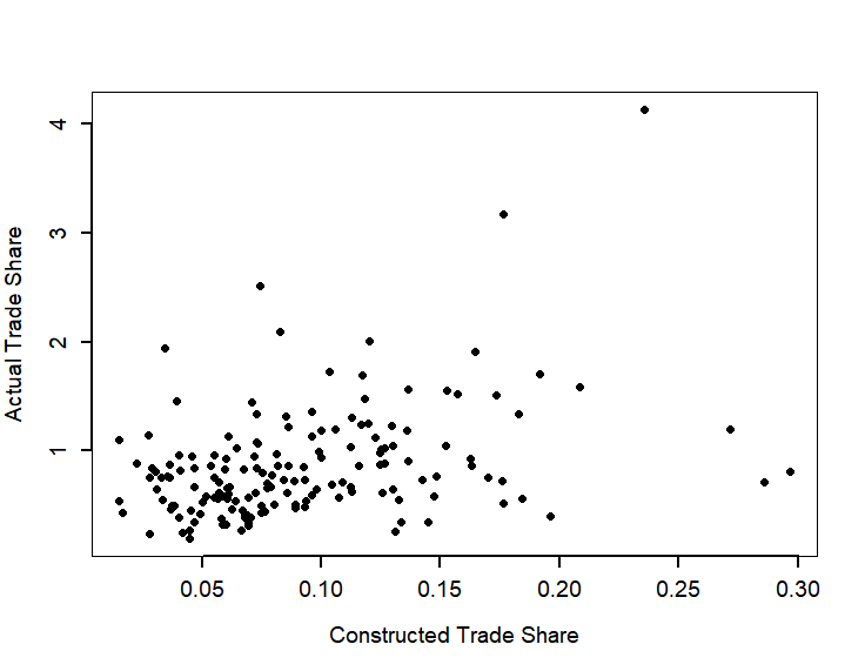}\\
  \caption{scatter plot of real and constructed trade share}\label{fig_emp1}
\end{figure}

The summary statistics of main data is presented in Table \ref{tab6}. Figure \ref{fig_emp1} is the scatter diagram of actual and constructed  share of international trade. Their correlation coefficient is 0.36.
\begin{table}[!htbp]
\centering
\caption{Summary statistics}
\label{tab6}
\begin{tabular}{ccccccc}
\hline
                        & mean   & std.dev  & median & minimum & maximum & sample size \\ \hline
Ln Income Per Capita    & 10.18  & 1.1      & 10.42  & 7.46    & 12.03   & 158         \\
Real Trade Share        & 0.87   & 0.52     & 0.76   & 0.2     & 4.13    & 158         \\
Constructed Trade Share & 0.09   & 0.05     & 0.08   & 0.02    & 0.3     & 158         \\
Ln Population           & 1.38   & 1.8      & 1.48   & -3.04   & 6.67    & 158         \\
Ln Area (Land)          & 11.73  & 2.26     & 12.02  & 5.7     & 16.61   & 158         \\
Area (Water)            & 25378  & 100818.4 & 2365   & 0       & 891163  & 158         \\
Coastline                & 4268.6 & 17451.71 & 523    & 0       & 202080  & 158         \\
Land Boundaries         & 2837.8 & 3407.8   & 1899.5 & 0       & 22147   & 158         \\
\% Forest               & 29.89  & 22.38    & 30.62  & 0       & 98.26   & 158         \\
\% Arable Land          & 40.95  & 21.55    & 42.06  & 0.56    & 82.56   & 158         \\
PM2.5                   & 25.05  & 19.43    & 22     & 5.9     & 100     & 158         \\
Languages               & 1.87   & 2.13     & 1      & 1       & 16      & 158         \\ \hline
\end{tabular}
\begin{tablenotes}
\item[1] NOTE: Income Per Capita is measured in dollars. Population is measured in millions. Land area and water area are measured in square kilometers. Coastline and land boundaries are measured in kilometers. PM2.5 is measured in micrograms per cubic meter. Source: \emph{FR99}, Penn World Table (PWT 9.1), the World Bank, and State of Global Air.
\end{tablenotes}
\end{table}

%which is consist of two parts. One is used to fit the optimal instruments, which include bilateral trade data, GDP, total population, capital stock, total land area, dummy variables for landlocked countries for each country, common border between two countries  and bilateral distance of two countries. The other is exogenous covariates to analyse the causal relationship between trade and growth, which include real trade share, total population, per capita income, total area, total land area, total water area, coastline, the arable land as percentage of total land, land boundaries, forest area as percentage of land area, the number of official and other commonly used languages in a country and dummy variables for landlocked countries.
\subsection{Empirical Results}
To investigate the influence of invalid instruments on the estimation of $\beta$, we first conduct the estimation using the same instruments as  \cite{Fan2018Nonparametric} and compare our estimated value with \emph{FR99} and NAIVE. Then we add another instrumental variable PM 2.5. The results are summarized in Table \ref{tab7} and \ref{tab8}. The first column is OLS estimator. The second estimator is the 2SLS estimator using the same instruments with \emph{FR99}. The third column is the NAIVE estimator and the last column is our estimator.
%the selected relevant instruments include the proxy for trade, the interaction term of proxy for trade and the arable land as percentage of total land and the interaction term of proxy for trade and forest area as percentage of land area by adaptive group Lasso with BIC. And the

\begin{figure}[htbp!]
  \centering
  % Requires \usepackage{graphicx}
  \includegraphics[width=14cm]{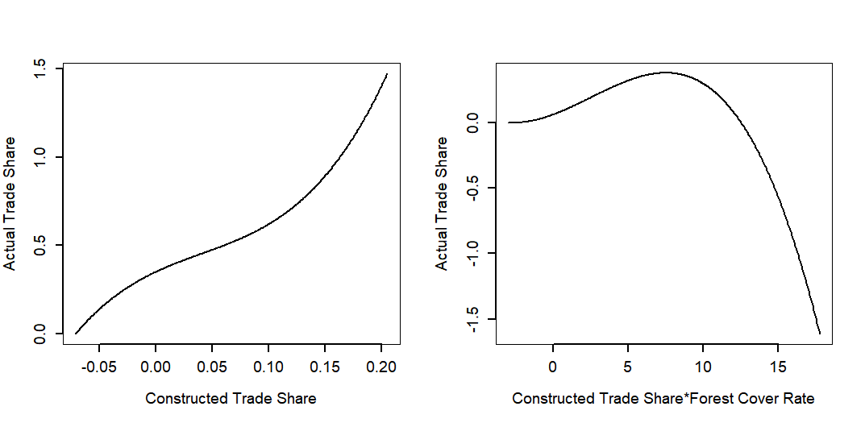}\\
  \caption{Plots of the endogenous variable (real trade share) against the selected four instrumental variables}\label{fig_emp2}
\end{figure}

When the variable PM2.5 is not included, the selected relevant instruments include the proxy for trade and the interaction term of proxy for trade and forest area as percentage of land area by the adaptive group Lasso with EBIC. The fitted functions of selected instruments by EBIC are plotted in Figure \ref{fig_emp2}. From Figure \ref{fig_emp2}, we see that proxy for trade and the interaction term of proxy for trade and forest area as percentage of land area instruments are likely to have nonlinear relationship with real trade share. In Table \ref{tab7}, the OLS estimator has severe bias and is inconsistent because of the endogeneity issue. The t statistics  for the NAIVE on trade is 3.85, compared to 2.99 for the FR99. As expected the each elements of $\boldsymbol{\alpha}$ is estimated to zero and our estimator is same as NAIVE. 
\begin{table}[htbp!]
\centering
\caption{Estimation results for the trade and income data (PM2.5 not included in the IV set)}
\label{tab7}
\begin{tabular}{ccccc}
\hline
                         & OLS       & 2SLS     & NAIVE    & R2IVE       \\ \hline
constant                 & 5.68E-08  & 4.93E-08 & 5.35E-08 & 5.35E-08 \\
                         & (0.08)     & (0.09)    & (0.08)    & (0.08)    \\
trade share              & 0.88***   & 1.43**   & 1.11***  & 1.11***  \\
                         & (0.18)     & (0.48)    & (0.29)    & (0.29)    \\
$R^{2}$                  & 0.13      & 0.05     & 0.07     & 0.07     \\
Sample Size              & 158       & 158      & 158      & 158      \\ \hline
\end{tabular}
\end{table}

\begin{table}[htbp!]
\centering
\caption{Estimation results for the trade and income data (PM2.5 included in the IV set)}
\label{tab8}
\begin{tabular}{ccccc}
\hline
                         & OLS       & 2SLS     & NAIVE    & R2IVE       \\ \hline
constant                 & 5.68E-08  & 4.93E-08 & 5.05E-08 & 5.52E-08 \\
                         & (0.08)     & (0.09)    & (0.08)    & (0.08)    \\
trade share              & 0.88***   & 1.43**   & 1.34***  & 1.19***  \\
                         & (0.18)     & (0.48)    & (0.26)    & (0.25)    \\
$R^{2} $                 & 0.13      & 0.05     & 0.15     & 0.13     \\
Sample Size              & 158       & 158      & 158      & 158      \\ \hline
\end{tabular}
\end{table}
When the variable PM 2.5 is included, the estimation results is summarized in Table \ref{tab8}. If we use NAIVE, under the operating assumption that all the instruments are valid, the estimated causal effect is 1.34 (with a standard error of 0.26). Our estimator can throw out the irrelevant instrument, the arable land as percentage of total land, and select the invalid instrument, PM2.5. Our estimator estimates the causal effect to be 1.19 which is close to the results in Table \ref{tab7}. This shows the R2IVE is robust to the potential invalid and irrelevant IVs. At last, compared with the original study of \emph{FR99}, the causal effect of trade on growth in 2010s is found to be smaller in magnitude but even more significant.

\section{Conclusion}

In this paper, we develop a robust IV estimator to both the invalid and irrelevant instruments (R2IVE) for the estimation of endogenous treatment effect, which extends the \emph{sisVIVE} \citep{Kang2015Instrumental} by considering a true high-dimensional instrumental variable setting and a general nonlinear reduced form equation. The proposed R2IVE  is shown to be root-$n$ consistent and asymptotically normal. Monte Carlo simulations demonstrate that the R2IVE performs better than the existing contemporary IV estimators (such as NAIVE and \emph{sisVIVE}) in many cases. The empirical study revisits the classic question of trade and growth. It is demonstrated that the R2IVE  can be applied to estimate the endogenous treatment effect with a large set of instruments without knowing which ones are relevant or valid and the reduced form is linear or nonlinear.
\newpage
\bibliographystyle{elsarticle-harv}
\bibliography{myref}

\section{Appendix}
Firstly, we need to clarify  the standard conditions for nonparametric estimation \citep{Huang2010} and adaptive Elastic-Net \citep{Zou2009Adenet}.
\begin{assumption}
(A1) The support of each instrument $Z_j$ is $[a,b]$,  where $a$ and $b$ are finite real numbers. And $Z_{ij}$ satisfies $\lim_{n\rightarrow \infty} \frac{\max_{i=1,...,n}\sum_{j=1}^{L}Z_{ij}^{2}}{n}=0$.  The density function $g_j$ of $Z_j$ in \eqref{eq4} satisfies $0 < C_1 < g_j(Z) < C_2 < \infty $ on $[a,b]$ for $j=1,...,L$. We use $\lambda_{\min}(\mathbf{M})$ and $\lambda_{\max}(\mathbf{M})$ to denote the minimum and maximum eigenvalues of a positive define matrix $\mathbf{M}$, respectively. Then we assume
\begin{equation*}
  b_1\leq\lambda_{\min}(\frac{1}{n}\mathbf{Z}^{T}\mathbf{Z})\leq \lambda_{\max}(\frac{1}{n}\mathbf{Z}^{T}\mathbf{Z})\leq B_1
\end{equation*}
where $b_1$ and $B_1$ are two positive constants.

(A2) Let $\mathcal{F}$ be the class of function $f$ such that the $k$th derivative $f^{(k)}$ exists and satisfies a Lipschitz condition of order $r \in (0,1]$. That is
$$\mathcal{F}=\left\{f(\cdot):\left|f^{(k)}(t_1)-f^{(k)}(t_2) \leq C\left|t_1-t_2\right|^{r}\right|,\text{ for } t_1, t_2 \in [a,b] \text{ and a constant } C > 0 \right\} $$
where $k$ is a nonnegative integer and $r \in (0,1]$ such that $s=k+r >1.5$. Suppose for  $f_{j} \in \mathcal{F}$, $j=1,...,L$ in \eqref{eq4}, there exists $C >0 $  such that $\min\limits_{j \in \mathcal{A}_{R}^{*}}\left\|f_{j}\right\| \geq C$, where $\left\|f_{j}\right\|_{2}^{2}=\int_{a}^{b}f_{j}^{2}(x)dx$.

(A3) 
$
\lim\limits_{n\rightarrow \infty}\frac{\log(L)}{\log(n)}=\eta \text{ for some } 0\leq\eta<1
$. $\lim\limits_{n\rightarrow \infty}\lambda_2/n=0$, $\lim\limits_{n\rightarrow \infty}\lambda_1/\sqrt{n}=0$ and $ \lim\limits_{n\rightarrow \infty}\lambda_1^{*}/\sqrt{n}=0$, $\lim\limits_{n\rightarrow \infty}\frac{\lambda_1^{*}}{\sqrt{n}}n^{\frac{(1-\eta)(1+\tau)-1}{2}}=\infty$. $\lim\limits_{n\rightarrow \infty}\frac{\lambda_2}{\sqrt{n}}\sqrt{\sum_{j \in \mathcal{A}_{I}}\alpha_{j}^{*2}} =0$ and $\lim\limits_{n\rightarrow \infty}\min(\frac{n}{\lambda_{1}\sqrt{p}},\left(\frac{\sqrt{n}} {\sqrt{p}\lambda_1^{*}}\right)^{\frac{1}{\tau}})(\min\limits_{ j \in \mathcal{A}_{I}}|\alpha_{j}^{*}|)\rightarrow \infty$.   $\tau = \left\lceil\frac{2\eta}{1-\eta}\right\rceil+1$. 
\end{assumption}
\subsection{Proof of Lemma 3.1}

 The \eqref{lem1}-\eqref{lem3} essentially follows the results of Theorem 3 in \cite{Huang2010}, we only give the proof of \eqref{lem4}.
\begin{equation}\label{pr2}
\begin{aligned}
\left\|\mathbf{D}^{*}-\widehat{\mathbf{D}}\right\| = & \left\|\sum_{j \in \mathcal{A}_{R}} f_{j} -\sum_{j \in \widehat{A}_{R}} \widehat{f}_{nj} \right\|\\
=&\left\|\sum_{j \in \widehat{A}_{R} \cap \mathcal{A}_{R}}\left(\widehat{f_{nj}}-f_{j}\right) -\sum_{j \in \widehat{A}_{R} \cap \mathcal{A}_{R}^{c}} \widehat{f_{nj}}+\sum_{j \in \widehat{\mathcal{A}}_{R}^{c} \cap \mathcal{A}_{R}} f_{j}\right\|\\
\leq &
\left\|\sum_{j \in \widehat{A}_{R} \cap \mathcal{A}_{R}}\left(\widehat{f_{nj}}-f_{j}\right)\right\| +\left\|\sum_{j \in \widehat{A}_{R} \cap \mathcal{A}_{R}^{c}} \widehat{f_{nj}}\left(z_{i j}\right)\right\|+\left\|\sum_{j \in \widehat{\mathcal{A}}_{R}^{c} \cap \mathcal{A}_{R}} f_{j}\left(z_{i j}\right)\right\|\\
\leq &
\sqrt{\sum_{j \in \mathcal{A}_{R}}\left\|\widehat{f_{n j}}-f_{j}\right\|_2^2} +\left\|\sum_{j \in \widehat{\mathcal{A}}_{R} \cap \mathcal{A}_{R}^{c}} \widehat{f_{nj}}\right\|+\left\|\sum_{j \in \widehat{\mathcal{A}}_{R}^{c} \cap \mathcal{A}_{R}} f_{j}\right\| \\
=&o_p\left(1\right)
\end{aligned}
\end{equation}
where the first term follows equation \eqref{lem3}  and the last two terms $\rightarrow 0$ because the selection consistency of adaptive group Lasso, that is \eqref{lem1}.
${\tiny\square}$
\subsection{Proof of Lemma 3.2}
We firstly show the following two equations hold and then the results of Lemma 3.2 holds by \cite{Zou2009Adenet}, Theorem 3.3.
\begin{align}
\left\|\mathcal{M}_{\widehat{\mathbf{D}}}\mathbf{Y}-\mathcal{M}_{\mathbf{D}^{*}}\mathbf{Y}\right\|&/n=o_p(1)\label{pr3}\\
\left\|\mathcal{M}_{\widehat{\mathbf{D}}}\mathbf{Z}-\mathcal{M}_{\mathbf{D}^{*}}\mathbf{Z}\right\|&/n=o_p(1)\label{pr4}
\end{align}
Before the  proof of \eqref{pr3} and \eqref{pr4}, we first prove the following conclusion.
\begin{equation}\label{pr3_1}
\left\|\mathcal{P}_{\mathbf{D}^{*}}-\mathcal{P}_{\widehat{\mathbf{D}}}\right\|=o_p(1)
\end{equation}

\begin{equation}\label{pr3_2}
\begin{aligned}
\left\|\mathcal{P}_{\mathbf{D}^{*}}-\mathcal{P}_{\widehat{\mathbf{D}}}\right\|
=&\left\|\mathbf{D}^{*}(\mathbf{D}^{*'}\mathbf{D}^{*})^{-1}\mathbf{D}^{*'} - \widehat{\mathbf{D}}(\widehat{\mathbf{D}}'\widehat{\mathbf{D}})^{-1}\widehat{\mathbf{D}}'\right\|\\
=&\left\|\mathbf{D}^{*}(\mathbf{D}^{*'}\mathbf{D}^{*})^{-1}\mathbf{D}^{*'} - (\widehat{\mathbf{D}}-\mathbf{D}^{*}+\mathbf{D}^{*})(\widehat{\mathbf{D}}'\widehat{\mathbf{D}})^{-1}(\widehat{\mathbf{D}}-\mathbf{D}^{*}+\mathbf{D}^{*})'\right\|\\
=&
\left\|\mathbf{D}^{*}\left[(\mathbf{D}^{*'}\mathbf{D}^{*})^{-1}-(\widehat{\mathbf{D}}'\widehat{\mathbf{D}})^{-1}\right]\mathbf{D}^{*'}
-(\widehat{\mathbf{D}}-\mathbf{D}^{*})(\widehat{\mathbf{D}}'\widehat{\mathbf{D}})^{-1}(\widehat{\mathbf{D}} - \mathbf{D}^{*})'-2(\widehat{\mathbf{D}}-\mathbf{D}^{*})(\widehat{\mathbf{D}}'\widehat{\mathbf{D}})^{-1}\mathbf{D}^{*'} \right\| \\
\leq & \left|(\mathbf{D}^{*'}\mathbf{D}^{*})^{-1}-(\widehat{\mathbf{D}}'\widehat{\mathbf{D}})^{-1}\right|\left\|\mathbf{D}^{*} \right\|^2+\left|(\widehat{\mathbf{D}}'\widehat{\mathbf{D}})^{-1}\right|\left\|\widehat{\mathbf{D}}-\mathbf{D}^{*}\right\|^2+2\left|(\widehat{\mathbf{D}}'\widehat{\mathbf{D}})^{-1}\right|\left\|\widehat{\mathbf{D}}-\mathbf{D}^{*}\right\|\left\|\mathbf{D}^{*}\right\|\\
=&\frac{\left|\left\|\widehat{\mathbf{D}}\right\|^2-\left\|\mathbf{D}^{*}\right\|^2\right|}{\left\|\widehat{\mathbf{D}}\right\|^2}
+\left|(\widehat{\mathbf{D}}'\widehat{\mathbf{D}})^{-1}\right|\left\|\widehat{\mathbf{D}}-\mathbf{D}^{*}\right\|^2+2\left|(\widehat{\mathbf{D}}'\widehat{\mathbf{D}})^{-1}\right|\left\|\widehat{\mathbf{D}}-\mathbf{D}^{*}\right\|\left\|\mathbf{D}^{*}\right\|\\
=:& S_1+S_2+S_3
\end{aligned}
\end{equation}
where $\mathbf{D}^{*'}\mathbf{D}^{*}$ and $\widehat{\mathbf{D}}'\widehat{\mathbf{D}}$ are real numbers and the first inequality is derived by triangle inequality.
\begin{equation}\label{pr3_3}
\begin{aligned}
S_1=& \frac{\left|\left\|\widehat{\mathbf{D}}-\mathbf{D}^{*}+\mathbf{D}^{*}\right\|^2-\left\|\mathbf{D}^{*}\right\|^2\right|}{\left\|\widehat{\mathbf{D}}\right\|^2}\\
= & \frac{\left\|\widehat{\mathbf{D}}-\mathbf{D}^{*}\right\|^2+2\left|\left(\widehat{\mathbf{D}}-\mathbf{D}^{*}\right)'\mathbf{D}^{*}\right|}{\left\|\widehat{\mathbf{D}}\right\|^2}\\
%\leq & \frac{\left\|\widehat{\mathbf{D}}-\mathbf{D}^{*}\right\|^2}{\left\|\widehat{\mathbf{D}}\right\|^2}
%+ %2\frac{\left\|\widehat{\mathbf{D}}-\mathbf{D}^{*}\right\|\left\|\mathbf{D}^{*}\right\|}{\left\|\widehat{\mathbf{D}}\right\|^2}\\
\leq & \frac{\left\|\widehat{\mathbf{D}}-\mathbf{D}^{*}\right\|^2}{n} \frac{1}{\left\|\widehat{\mathbf{D}}\right\|^2/n}
+2 \frac{\left\|\widehat{\mathbf{D}}-\mathbf{D}^{*}\right\|}{\sqrt{n}}
\frac{\left\|\mathbf{D}^{*}\right\|/\sqrt{n}}{\left\|\widehat{\mathbf{D}}\right\|^2/n}\\
=&o_p(1)O_p(1)+o_p(1)O_p(1)\\
=&o_p(1)
\end{aligned}
\end{equation}
\begin{equation}\label{pr3_4}
S_2= \frac{\left\|\widehat{\mathbf{D}}-\mathbf{D}^{*}\right\|^2}{n} \frac{1}{\left\|\widehat{\mathbf{D}}\right\|^2/n}=o_p(1)O_p(1)=o_p(1)
\end{equation}
\begin{equation}\label{pr3_5}
S_3= \frac{\left\|\widehat{\mathbf{D}}-\mathbf{D}^{*}\right\|}{\sqrt{n}}
\frac{\left\|\mathbf{D}^{*}\right\|/\sqrt{n}}{\left\|\widehat{\mathbf{D}}\right\|^2/n}=o_p(1)O_p(1)=o_p(1)
\end{equation}
Substituting \eqref{pr3_3}-\eqref{pr3_5} into \eqref{pr3_2}, we have $\left\|\mathcal{P}_{\mathbf{D}^{*}}-\mathcal{P}_{\widehat{\mathbf{D}}}\right\|=o_p(1)$. After the proof of \eqref{pr3_1}, we now focus on the proof of \eqref{pr3}.
We present the matrix form of \eqref{eq12}
\begin{equation}\label{pr5}
\mathbf{Y}=\mathbf{D}^{*}\beta^{*}+\mathbf{Z}\boldsymbol{\alpha}^{*}+\boldsymbol{\nu}
\end{equation}
Substituting \eqref{pr5} into \eqref{pr3}, we have
\begin{equation}\label{pr3_6}
\begin{aligned}
\left\|\mathcal{M}_{\widehat{\mathbf{D}}}\mathbf{Y}-\mathcal{M}_{\mathbf{D}^{*}}\mathbf{Y}\right\|/n
=&\left\|\left(\mathcal{P}_{\mathbf{D}^{*}}-\mathcal{P}_{\widehat{\mathbf{D}}}\right)\mathbf{Y}\right\|/n\\
=&\left\|\left(\mathcal{P}_{\mathbf{D}^{*}}-\mathcal{P}_{\widehat{\mathbf{D}}}\right)\left(\mathbf{D}^{*}\beta^{*}+\mathbf{Z}\boldsymbol{\alpha}^{*}+\boldsymbol{\nu}\right)\right\|/n\\
\leq &\left\|\left(\mathcal{P}_{\mathbf{D}^{*}}-\mathcal{P}_{\widehat{\mathbf{D}}}\right)\mathbf{D}^{*}\beta^{*}\right\|/n
+\left\|\left(\mathcal{P}_{\mathbf{D}^{*}}-\mathcal{P}_{\widehat{\mathbf{D}}}\right)\mathbf{Z}\boldsymbol{\alpha}^{*}\right\|/n
+\left\|\left(\mathcal{P}_{\mathbf{D}^{*}}-\mathcal{P}_{\widehat{\mathbf{D}}}\right)\boldsymbol{\nu}\right\|/n\\
\leq &\left\|\mathcal{P}_{\mathbf{D}^{*}}-\mathcal{P}_{\widehat{\mathbf{D}}}\right\|\frac{\left\|\mathbf{D}^{*}\right\|}{n}\left|\beta^{*}\right|
+\left\|\mathcal{P}_{\mathbf{D}^{*}}-\mathcal{P}_{\widehat{\mathbf{D}}}\right\|\frac{\left\|\mathbf{Z}\boldsymbol{\alpha}^{*}\right\|}{n}
+\left\|\mathcal{P}_{\mathbf{D}^{*}}-\mathcal{P}_{\widehat{\mathbf{D}}}\right\|\frac{\left\|\boldsymbol{\nu}\right\|}{n}\\
=&o_p(1)O_p(1)+o_p(1)O_p(1)+o_p(1)O_p(1)\\
=&o_p(1)
\end{aligned}
\end{equation}
Similarly,
\begin{equation}\label{pr3_7}
\begin{aligned}
\left\|\mathcal{M}_{\widehat{\mathbf{D}}}\mathbf{Z}-\mathcal{M}_{\mathbf{D}^{*}}\mathbf{Z}\right\|/n
=&\left\|\left(\mathcal{P}_{\mathbf{D}^{*}}-\mathcal{P}_{\widehat{\mathbf{D}}}\right)\mathbf{Z}\right\|/n\\
\leq & \left\|\mathcal{P}_{\mathbf{D}^{*}}-\mathcal{P}_{\widehat{\mathbf{D}}}\right\|\left\|\mathbf{Z}\right\|/n\\
= & o_p(1)O_p(1)\\
=&o_p(1)
\end{aligned}
\end{equation}

${\tiny\square}$
\subsection{Proof of Theorem 3.1}

Substituting \eqref{pr5} into \eqref{estimator}, we have
\begin{equation}\label{pr6}
\begin{aligned}
\widehat{\beta}& =\left(\widehat{\mathbf{D}}'\mathcal{M}_{\widehat{\mathcal{A}}_{I}}\widehat{\mathbf{D}}\right)^{-1} \widehat{\mathbf{D}}'\mathcal{M}_{\widehat{\mathcal{A}}_{I}}\mathbf{Y}\\
& = \left(\widehat{\mathbf{D}}'\mathcal{M}_{\widehat{\mathcal{A}}_{I}}\widehat{\mathbf{D}}\right)^{-1} \widehat{\mathbf{D}}'\mathcal{M}_{\widehat{\mathcal{A}}_{I}} \left(\mathbf{D}^{*}\beta^{*}+\mathbf{Z}\boldsymbol{\alpha}^{*}+\boldsymbol{\nu}\right)\\ & = \left(\widehat{\mathbf{D}}'\mathcal{M}_{\widehat{\mathcal{A}}_{I}}\widehat{\mathbf{D}}\right)^{-1} \widehat{\mathbf{D}}'\mathcal{M}_{\widehat{\mathcal{A}}_{I}}\left[\left(\widehat{\mathbf{D}}+\mathbf{D}^{*} -\widehat{\mathbf{D}}\right)\beta^{*}+\mathbf{Z}\boldsymbol{\alpha}^{*}+\boldsymbol{\nu}\right]\\
& = \beta^{*}+ \left(\widehat{\mathbf{D}}'\mathcal{M}_{\widehat{\mathcal{A}}_{I}}\widehat{\mathbf{D}}\right)^{-1} \widehat{\mathbf{D}}'\mathcal{M}_{\widehat{\mathcal{A}}_{I}} \left[\left(\mathbf{D}^{*}-\widehat{\mathbf{D}}\right)\beta^{*}+\mathbf{Z}\boldsymbol{\alpha}^{*}+\boldsymbol{\nu}\right]\\
\end{aligned}
\end{equation}
which yields that
\begin{equation}\label{pr7}
\begin{aligned}
\sqrt{n}\left(\widehat{\beta}-\beta^{*}\right)& = \left(\frac{\widehat{\mathbf{D}}'\mathcal{M}_{\widehat{\mathcal{A}}_{I}}\widehat{\mathbf{D}}}{n}\right)^{-1}
\frac{\widehat{\mathbf{D}}'\mathcal{M}_{\widehat{\mathcal{A}}_{I}} \left[\left(\mathbf{D}^{*}-\widehat{\mathbf{D}}\right)\beta^{*}+\mathbf{Z}\boldsymbol{\alpha}^{*}+\boldsymbol{\nu}\right]}{\sqrt{n}}\\
& =: T_1^{-1} +T_2
\end{aligned}
\end{equation}
We want to show
\begin{equation}\label{pr8}
\sqrt{n}\left(\widehat{\beta}-\beta^{*}\right)= \left(\frac{\mathbf{D}^{*'} \mathcal{M}_{\mathcal{A}_{I}}\mathbf{D}^{*}}{n}+o_p(1)\right)^{-1} \left(\frac{\mathbf{D}^{*'}\mathcal{M}_{\mathcal{A}_{I}}\boldsymbol{\nu}}{\sqrt{n}}+o_p(1)\right)
\end{equation}

Before we show \eqref{pr8}, we first prove the following conclusions.
\begin{equation}\label{pr10}
\left\|\mathcal{M}_{\widehat{\mathcal{A}}_{I}}\left(\widehat{\mathbf{D}}-\mathbf{D}^{*}\right)\right\|= o_p(1)
\end{equation}
\begin{equation} \label{pr15}
\begin{aligned}
\left\|\mathcal{M}_{\widehat{\mathcal{A}}_{I}}\left(\widehat{\mathbf{D}}-\mathbf{D}^{*}\right)\right\| & \leq \left\|\widehat{\mathbf{D}}-\mathbf{D}^{*}\right\| + \left\|\mathcal{P}_{\widehat{\mathcal{A}}_{I}}\left(\widehat{\mathbf{D}}-\mathbf{D}^{*}\right)\right\|
\end{aligned}
\end{equation}
where the second term is
\begin{equation} \label{pr16}
\begin{aligned}
\left\|\mathcal{P}_{\widehat{\mathcal{A}}_{I}}\left(\widehat{\mathbf{D}}-\mathbf{D}^{*}\right)\right\| & \leq \left\|\mathcal{P}_{\widehat{\mathcal{A}}_{I}}\right\|\left\|\widehat{\mathbf{D}}-\mathbf{D}^{*}\right\|\\ & =  \left\|\widehat{\mathbf{D}}-\mathbf{D}^{*}\right\|\\
%& \leq B_2/b_2\left\|\widehat{\mathbf{D}}-\mathbf{D}^{*}\right\|/\sqrt{n}\\
& =o_p(1)
\end{aligned}
\end{equation}
where the first inequality follows Cauchy-Schwarz inequality, the second step is because that the largest eigenvalues of projection matrix is 1 and the last step is from the equation \eqref{lem4}.
Substituting \eqref{pr16} into \eqref{pr15} combining with \eqref{lem4}, we have $\left\|\mathcal{M}_{\widehat{\mathcal{A}}_{I}}\left(\widehat{\mathbf{D}}-\mathbf{D}^{*}\right)\right\|= o_p(1)$.

After the proof of \eqref{pr10}, we now go back to \eqref{pr8}. We first focus on the term $T_1$.
\begin{equation}\label{pr18}
\begin{aligned}
T_1 &= \frac{\widehat{\mathbf{D}}'\mathcal{M}_{\widehat{\mathcal{A}}_{I}}\widehat{\mathbf{D}}}{n}\\
& = \frac{\left(\widehat{\mathbf{D}}-\mathbf{D}^{*}+\mathbf{D}^{*}\right)'\mathcal{M}_{\widehat{\mathcal{A}}_{I}}\left(\widehat{\mathbf{D}}-\mathbf{D}^{*}+\mathbf{D}^{*}\right)}{n}\\
& = \frac{\mathbf{D}^{*'} \mathcal{M}_{\mathcal{A}_{I}}\mathbf{D}^{*}}{n}
+  \frac{\left(\widehat{\mathbf{D}}-\mathbf{D}^{*}\right)'\mathcal{M}_{\widehat{\mathcal{A}}_{I}}\left(\widehat{\mathbf{D}}-\mathbf{D}^{*}\right)}{n}
+
2\frac{\mathbf{D}^{*'}\mathcal{M}_{\widehat{\mathcal{A}}_{I}}\left(\widehat{\mathbf{D}}-\mathbf{D}^{*}\right)}{n}
+ \frac{\mathbf{D}^{*'}\left(\mathcal{M}_{\widehat{\mathcal{A}}_{I}}-\mathcal{M}_{\mathcal{A}_{I}}\right)\mathbf{D}^{*}}{n}\\
& =: \frac{\mathbf{D}^{*'} \mathcal{M}_{\mathcal{A}_{I}}\mathbf{D}^{*}}{n} +T_{11}+2\cdot T_{12}+T_{13}
\end{aligned}
\end{equation}
For the term $T_{11}$, we have
\begin{equation}\label{pr19}
\begin{aligned}
\left|T_{11}\right| & = \left|\left(\widehat{\mathbf{D}}-\mathbf{D}^{*}\right)'\mathcal{M}_{\widehat{\mathcal{A}}_{I}}\left(\widehat{\mathbf{D}}-\mathbf{D}^{*}\right)/n\right|\\
& \leq \left\|\mathcal{M}_{\widehat{\mathcal{A}}_{I}}\left(\widehat{\mathbf{D}}-\mathbf{D}^{*}\right)/\sqrt{n}\right\| \left\|\mathcal{M}_{\widehat{\mathcal{A}}_{I}}\left(\widehat{\mathbf{D}}-\mathbf{D}^{*}\right)/\sqrt{n}\right\|\\
& =o_p(1)
\end{aligned}
\end{equation}
where the first inequality holds by Cauchy-Schwarz inequality and the last step is derived by \eqref{pr10}. Similarly, we deal with the  term $T_{12}$,

\emph{\begin{equation}\label{pr20}
\begin{aligned}
\left|T_{12}\right| & =
\left|\mathbf{D}^{*'}\mathcal{M}_{\widehat{\mathcal{A}}_{I}}\left(\widehat{\mathbf{D}}-\mathbf{D}^{*}\right)/n\right|\\
& \leq
\left\|\mathbf{D}^{*}/\sqrt{n}\right\| \left\|\mathcal{M}_{\widehat{\mathcal{A}}_{I}}\left(\widehat{\mathbf{D}}-\mathbf{D}^{*}\right)/\sqrt{n}\right\|\\
& \leq
\left\|\mathbf{D}/\sqrt{n}\right\| \left\|\mathcal{M}_{\widehat{\mathcal{A}}_{I}}\left(\widehat{\mathbf{D}}-\mathbf{D}^{*}\right)/\sqrt{n}\right\|\\
&= \sqrt{\sum_{i=1}^{n}D_i^2/n}\left\|\mathcal{M}_{\widehat{\mathcal{A}}_{I}}\left(\widehat{\mathbf{D}}-\mathbf{D}^{*}\right)/\sqrt{n}\right\|\\
&=([E(D_i^2)]^{1/2}+o_p(1))o_p(1)\\
&= o_p(1)
\end{aligned}
\end{equation}}
Next, we deal with the term $T_{13}$
\begin{equation} \label{pr21}
\begin{aligned}
T_{13}=\frac{\mathbf{D}^{*'}\left(\mathcal{M}_{\widehat{\mathcal{A}}_{I}}-\mathcal{M}_{\mathcal{A}_{I}}\right)\mathbf{D}^{*}}{n}=o_p(1)
\end{aligned}
\end{equation}
which holds since $P(\widehat{\mathcal{A}}_{I}=\mathcal{A}_{I})\rightarrow 1$.
Substituting \eqref{pr19}, \eqref{pr20} and \eqref{pr21} into \eqref{pr18}, we have
\begin{equation}\label{pr22}
\begin{aligned}
T_1  = \frac{\mathbf{D}^{*'} \mathcal{M}_{\mathcal{A}_{I}}\mathbf{D}^{*}}{n} +o_p(1)
\end{aligned}
\end{equation}

Now, we work on the term $T_2$.
\begin{equation}\label{pr23}
\begin{aligned}
T_2 = &\frac{\widehat{\mathbf{D}}'\mathcal{M}_{\widehat{\mathcal{A}}_{I}}\left[\left(\mathbf{D}^{*}-\widehat{\mathbf{D}}\right)\beta^{*} +\mathbf{Z}\boldsymbol{\alpha}^{*}+\boldsymbol{\nu}\right]}{\sqrt{n}}\\
= & \frac{\left(\widehat{\mathbf{\mathbf{D}}}-\mathbf{D}^{*}+\mathbf{D}^{*}\right)'\mathcal{M}_{\widehat{\mathcal{A}}_{I}} \left[\left(\mathbf{D}^{*}-\widehat{\mathbf{D}}\right)\beta^{*}+\mathbf{Z}\boldsymbol{\alpha}^{*}+\boldsymbol{\nu}\right]}{\sqrt{n}}\\
= &
\frac{\mathbf{D}^{*'}\mathcal{M}_{\mathcal{A}_{I}}\boldsymbol{\nu}}{\sqrt{n}}
-\frac{\left(\widehat{\mathbf{\mathbf{D}}}-\mathbf{D}^{*}\right)'\mathcal{M}_{\widehat{\mathcal{A}}_{I}}\left(\widehat{\mathbf{\mathbf{D}}}-\mathbf{D}^{*}\right)\beta^{*}}{\sqrt{n}}
+\frac{\left(\widehat{\mathbf{\mathbf{D}}}-\mathbf{D}^{*}\right)'\mathcal{M}_{\widehat{\mathcal{A}}_{I}}\mathbf{Z}\boldsymbol{\alpha}^{*}}{\sqrt{n}}\\
&
+\frac{\left(\widehat{\mathbf{\mathbf{D}}}-\mathbf{D}^{*}\right)'\mathcal{M}_{\widehat{\mathcal{A}}_{I}}\boldsymbol{\nu}}{\sqrt{n}}
-\frac{\mathbf{D}^{*'}\mathcal{M}_{\widehat{\mathcal{A}}_{I}}\left(\widehat{\mathbf{\mathbf{D}}}-\mathbf{D}^{*}\right)\beta^{*}}{\sqrt{n}}
+\frac{\mathbf{D}^{*'}\mathcal{M}_{\widehat{\mathcal{A}}_{I}}\mathbf{Z}\boldsymbol{\alpha}^{*}}{\sqrt{n}}
+\frac{\mathbf{D}^{*'}\left(\mathcal{M}_{\widehat{\mathcal{A}}_{I}}-\mathcal{M}_{\mathcal{A}_{I}}\right)\boldsymbol{\nu}}{\sqrt{n}}\\
=: & \frac{\mathbf{D}^{*'}\mathcal{M}_{\mathcal{A}_{I}}\boldsymbol{\nu}}{\sqrt{n}} -T_{21}+T_{22}+T_{23}-T_{24}+T_{25}+T_{26}
\end{aligned}
\end{equation}
Similar to \eqref{pr19}, we have
\begin{equation}\label{pr24}
\begin{aligned}
\left|T_{21}\right| &= \left|\left(\widehat{\mathbf{\mathbf{D}}}-\mathbf{D}^{*}\right)'\mathcal{M}_{\widehat{\mathcal{A}}_{I}}\left(\widehat{\mathbf{\mathbf{D}}}-\mathbf{D}^{*}\right)\beta^{*}/\sqrt{n}\right|\\
& \leq \left\|\mathcal{M}_{\widehat{\mathcal{A}}_{I}}\left(\widehat{\mathbf{\mathbf{D}}}-\mathbf{D}^{*}\right)\right\| \left\|\mathcal{M}_{\widehat{\mathcal{A}}_{I}}\left(\widehat{\mathbf{\mathbf{D}}}-\mathbf{D}^{*}\right)\right\| \left|\beta^{*}\right|/\sqrt{n}\\
& = o_p(1)
\end{aligned}
\end{equation}
For the second term of $T_{21}$, we have
\begin{equation}\label{pr25}
\begin{aligned}
\left|T_{22}\right|&=\left|\left(\widehat{\mathbf{\mathbf{D}}}-\mathbf{D}^{*}\right)'\mathcal{M}_{\widehat{\mathcal{A}}_{I}}\mathbf{Z}\boldsymbol{\alpha}^{*}/\sqrt{n}\right|\\
\leq &   \left\|\mathcal{M}_{\widehat{\mathcal{A}}_{I}}\left(\widehat{\mathbf{\mathbf{D}}}-\mathbf{D}^{*}\right)\right\| \left\|\mathcal{M}_{\widehat{\mathcal{A}}_{I}}\mathbf{Z}\boldsymbol{\alpha}^{*}/\sqrt{n}\right\|\\
= &  \left\|\mathcal{M}_{\widehat{\mathcal{A}}_{I}}\left(\widehat{\mathbf{\mathbf{D}}}-\mathbf{D}^{*}\right)\right\| \left\|(\mathcal{M}_{\widehat{\mathcal{A}}_{I}}-\mathcal{M}_{\mathcal{A}_{I}}+\mathcal{M}_{\mathcal{A}_{I}})\mathbf{Z}\boldsymbol{\alpha}^{*}/\sqrt{n}\right\|\\
\leq &  \left\|\mathcal{M}_{\widehat{\mathcal{A}}_{I}}\left(\widehat{\mathbf{\mathbf{D}}}-\mathbf{D}^{*}\right)\right\| \left\|(\mathcal{M}_{\widehat{\mathcal{A}}_{I}}-\mathcal{M}_{\mathcal{A}_{I}})\mathbf{Z}\boldsymbol{\alpha}^{*}/\sqrt{n}\right\|\\
&+ \left\|\mathcal{M}_{\widehat{\mathcal{A}}_{I}}\left(\widehat{\mathbf{\mathbf{D}}}-\mathbf{D}^{*}\right)/\sqrt{n}\right\| \left\|\mathcal{M}_{\mathcal{A}_{I}}\mathbf{Z}\boldsymbol{\alpha}^{*}/\sqrt{n}\right\|\\
= & o_p(1)+0\\
= & o_p(1)
\end{aligned}
\end{equation}
where for the terms in the last inequality, the first term is $o_p(1)$ as derived by the equation \eqref{pr10} and the consistency of variable selection. The second term is equal to 0 since $\mathcal{M}_{\mathcal{A}_{I}}\mathbf{Z}\boldsymbol{\alpha}^{*}=0$. 

Then, we deal with term $T_{23}$:

\begin{equation}\label{pr26}
\begin{aligned}
\left|T_{23}\right|&=\left|\left(\widehat{\mathbf{\mathbf{D}}}-\mathbf{D}^{*}\right)'\mathcal{M}_{\widehat{\mathcal{A}}_{I}}\boldsymbol{\nu}/\sqrt{n}\right|\\
& \leq \left\|\mathcal{M}_{\widehat{\mathcal{A}}_{I}}\left(\widehat{\mathbf{\mathbf{D}}}-\mathbf{D}^{*}\right)\right\|\left\|\boldsymbol{\nu}/\sqrt{n}\right\|\\
&=o_p(1)
\end{aligned}
\end{equation}
Similar to \eqref{pr20}, we have
\begin{equation}\label{pr27}
\begin{aligned}
\left|T_{24}\right|&=\left|\mathbf{D}^{*'}\mathcal{M}_{\widehat{\mathcal{A}}_{I}}\left(\widehat{\mathbf{\mathbf{D}}}-\mathbf{D}^{*}\right)\beta^{*}/\sqrt{n}\right|\\
& \leq \left\|\frac{\mathbf{D}^{*}}{\sqrt{n}}\right\| \left\|\mathcal{M}_{\widehat{\mathcal{A}}_{I}}\left(\widehat{\mathbf{\mathbf{D}}}-\mathbf{D}^{*}\right)\right\| \left|\beta^{*}\right|\\
&=o_p(1)
\end{aligned}
\end{equation}

Next, we deal with the term $T_{25}$
\begin{equation}\label{pr28}
\begin{aligned}
\left|T_{25}\right|&=\left|\frac{\mathbf{D}^{*'}\mathcal{M}_{\widehat{\mathcal{A}}_{I}}\mathbf{Z}\boldsymbol{\alpha}^{*}}{\sqrt{n}}\right|\\
&=\left|\frac{\mathbf{D}^{*'}\left(\mathcal{M}_{\widehat{\mathcal{A}}_{I}}-\mathcal{M}_{\mathcal{A}_{I}}+\mathcal{M}_{\mathcal{A}_{I}}\right)\mathbf{Z}\boldsymbol{\alpha}^{*}}{\sqrt{n}}\right|\\
& \leq \left|\frac{\mathbf{D}^{*'}\left(\mathcal{M}_{\widehat{\mathcal{A}}_{I}}-\mathcal{M}_{\mathcal{A}_{I}}\right)\mathbf{Z}\boldsymbol{\alpha}^{*}}{\sqrt{n}}\right|
+\left|\frac{\mathbf{D}^{*'}\mathcal{M}_{\mathcal{A}_{I}}\mathbf{Z}\boldsymbol{\alpha}^{*}}{\sqrt{n}}\right|\\
& \leq \left|\frac{\mathbf{D}^{*'}\left(\mathcal{M}_{\widehat{\mathcal{A}}_{I}}-\mathcal{M}_{\mathcal{A}_{I}}\right)\mathbf{Z}\boldsymbol{\alpha}^{*}}{\sqrt{n}}\right|
 + \left\|\frac{\mathbf{D}^{*}}{\sqrt{n}}\right\| \left\|\mathcal{M}_{\mathcal{A}_{I}}\mathbf{Z}\boldsymbol{\alpha}^{*}\right\|\\
& =o_p(1)+0=o_p(1)
\end{aligned}
\end{equation}
where for the terms in the last inequality, the first term is $o_p(1)$ following the variables selection consistency in Lemma 3.2, the second term is equal to 0.

Similar to \eqref{pr21}, we have
\begin{equation}\label{pr29}
\begin{aligned}
T_{26}=\frac{\mathbf{D}^{*'}\left(\mathcal{M}_{\widehat{\mathcal{A}}_{I}}-\mathcal{M}_{\mathcal{A}_{I}}\right)\boldsymbol{\nu}}{\sqrt{n}}=o_p(1)
\end{aligned}
\end{equation}
Combining \eqref{pr23}-\eqref{pr29}, we have
\begin{equation}\label{pr30}
\begin{aligned}
T_2 = \frac{\mathbf{D}^{*'}\mathcal{M}_{\mathcal{A}_{I}}\boldsymbol{\nu}}{\sqrt{n}} +o_p(1)
\end{aligned}
\end{equation}
Combining \eqref{pr30} with \eqref{pr22} yields \eqref{pr8}.
\begin{equation*}
\sqrt{n}\left(\widehat{\beta}-\beta^{*}\right)= \left(\frac{\mathbf{D}^{*'} \mathcal{M}_{\mathcal{A}_{I}}\mathbf{D}^{*}}{n} +o_p(1)\right)^{-1}\left(\frac{\mathbf{D}^{*'}\mathcal{M}_{\mathcal{A}_{I}}\boldsymbol{\nu}}{\sqrt{n}}+o_p(1)\right)
\end{equation*}

Because $\mathbf{D}^{*'} \mathcal{M}_{\mathcal{A}_{I}}\mathbf{D}^{*}/n= E\left(\mathbf{D}^{*'} \mathcal{M}_{\mathcal{A}_{I}}\mathbf{D}^{*}\right)+o_p(1)$ by the Weak Law of Large Numbers, note that $\sum_{i=1}^{n}D_i^{*}\mathcal{M}_{\mathcal{A}_{I}(i,j)}\nu_{j}$ are independent and identically distributed with mean zero and variance $\sigma_{n}^{2}=\left[E\left(\mathbf{D}^{*'} \mathcal{M}_{\mathcal{A}_{I}}\mathbf{D}^{*}\right)\right]^{-1}E\left(\mathbf{D}^{*'} \mathcal{M}_{\mathcal{A}_{I}}\mathbf{D}^{*}\nu_{i}^{2}\right)\left[E\left(\mathbf{D}^{*'} \mathcal{M}_{\mathcal{A}_{I}}\mathbf{D}^{*}\right)\right]^{-1}$ the Central Limit Theorem imply that
\begin{equation}\label{pr31}
 \sigma_{n}^{-1}\sqrt{n}\left(\widehat{\beta}-\beta^{*}\right)\rightarrow N(0,1)
\end{equation}
Under homoscedastic case, we have $\sigma_{n}^{2}=\left(E\left(\mathbf{D}^{*'} \mathcal{M}_{\mathcal{A}_{I}}\mathbf{D}^{*}\right)\right)^{-1}\sigma_{\boldsymbol{\nu}}^{2}$ with Var$(\nu_{i})=\sigma_{\boldsymbol{\nu}}^{2}$.
${\tiny\square}$

\end{document}